\documentclass[preprint,11pt,3p]{elsarticle}

\usepackage{amssymb}
\usepackage{graphicx,multirow,caption,subcaption,multicol,setspace,amsmath,wrapfig}
\usepackage[linesnumbered,ruled]{algorithm2e}
\usepackage{algorithmic}
\usepackage{adjustbox}
\usepackage{booktabs,caption}
\usepackage[flushleft]{threeparttable}
\usepackage{lineno}
\usepackage{enumitem}
\setlist{nosep}
\usepackage{rotating,caption}
\usepackage[table]{xcolor}
\usepackage{soul,xcolor}
\usepackage{amsfonts}
\usepackage{makecell,eqnarray}
\biboptions{numbers,sort&compress}
\newcommand{\cmark}{\ding{51}}%
\newcommand{\xmark}{\ding{55}}%
\graphicspath{ {Images/} }

\makeatletter
\newcommand{\algorithmfootnote}[2][\footnotesize]{%
  \let\old@algocf@finish\@algocf@finish
  \def\@algocf@finish{\old@algocf@finish
    \leavevmode\rlap{\begin{minipage}{\linewidth}
    #1#2
    \end{minipage}}%
  }%
}

\journal{Neurocomputing}

\begin{document}
\begin{frontmatter}



\title{Two-Dimensional (2D) Particle Swarms for Structure Selection of Nonlinear Systems}

\author[label5]{Faizal Hafiz\corref{cor1}}
\address[label5]{Department of Electrical \& Computer Engineering, The University of Auckland, Auckland, New Zealand}
\ead{faizalhafiz@ieee.org}
\cortext[cor1]{Corresponding author}

\author[label5]{Akshya Swain}
\ead{a.swain@auckland.ac.nz}

\author[label1]{Eduardo MAM Mendes}
\address[label1]{Department of Electronics Engineering, Federal University of Minas Gerais, Belo Horizonte, Brazil}

\begin{abstract}

The present study proposes a new structure selection approach for non-linear system identification based on Two-Dimensional particle swarms (2D-UPSO). The 2D learning framework essentially extends the learning dimension of the conventional particle swarms and explicitly incorporates the information about the cardinality, \textit{i.e.}, number of terms, into the search process. This property of the 2D-UPSO has been exploited to determine the correct structure of the non-linear systems. The efficacy of the proposed approach is demonstrated by considering several simulated benchmark nonlinear systems in discrete and in continuous domain. In addition, the proposed approach is applied to identify a parsimonious structure from practical non-linear wave-force data. The results of the comparative investigation with Genetic Algorithm (GA), Binary Particle Swarm Optimization (BPSO) and the classical Orthogonal Forward Regression (OFR) methods illustrate that the proposed 2D-UPSO could successfully detect the correct structure of the non-linear systems.

\end{abstract}


\end{frontmatter}

\section{Introduction}
\label{sec1:intro}

Construction of mathematical models from observed input-output data; popularly known as {\it system identification}, has been a major research concern from diverse fields such as statistics, control theory, information theory, economics, ecology, and agriculture. Most of the practical systems are inherently nonlinear and therefore the development of system identification methods, which can be valid for a broad class of nonlinear systems, has attracted many researchers~\citep{Verhaegen:2007,Billings:2013,Ljung:1999,Westwick:Verhaegen:1996}. 

In particular, this study aims to develop a new approach for the identification of nonlinear systems represented by polynomial Nonlinear Auto-Regressive with eXogenous inputs (NARX) models~\citep{Leontaritis:Billings:1985,LEONTARITIS:BILLINGS:1985b,Chen:Billings:1989b}. The NARX model enables the representation of a broad class of nonlinear systems in recursive lagged input and output terms which leads to a convenient \textit{linear-in-parameter} form. The major challenge of the non-linear system identification is to determine which input-output terms from the NARX model shall be included in the system model to capture its dynamics, referred to as the \textit{structure selection problem}. This difficulty can be ascribed to the exponential increase in the number of terms with the increase in the degree of non-linearity and maximum lags of inputs and outputs of the NARX model.

Over the past few decades, several search methods have been proposed to address the issue of nonlinear system identification, both, in time and frequency domain~\citep{Haber:Unbehauen:1990,Falsone:Piroddi:2015,Avellina:Piroddi:2017,Baldacchino:Kadirkamanathan:2013,Baldacchino:Kadirkamanathan:2012,Piroddi:Farina:2012,Bonin:Pirrodi:2010,Cantelmo:Piroddi:2010,Hong:Mitchell:2008,Billings:Wei:2008,Mendez:2001,Rodriguez:Fonseca:2004,Guo:Guo:2016,Tang:Long:2019,Vargas:Pedrycz:2019}. Most of these approaches can broadly be categorized into either \textit{sequential} or \textit{subset} search. This difference arises from the way the structure is built.

In the sequential search, a structure is built incrementally by adding and/or removing a term in each step. Among the sequential search methods, Orthogonal Forward Regression (OFR) with Error Reduction Ratio (ERR) has been extensively studied and applied to many applications by Billings and co-workers~\citep{Korenberg:Billings:1988,Chen:Billings:1989,Billings:Chen:Korenberg:1989,Chiras:Evans:2001,Hong:Mitchell:2008}. OFR is essentially a \textit{sequential greedy} search method, wherein the structure is built incrementally by including the term with the highest performance metric (\textit{i.e.}, ERR) in each step. The main limitation of OFR-ERR is the emphasis on a single term; though the terms are decoupled through orthogonalization, the performance metric such as ERR may depend on the order in which terms are orthogonalized~\citep{Mao:Billings:1997}. Several variants have been proposed to further improve the performance of OFR~\citep{Mendez:2001,Piroddi:Spinelli:2003,Billings:Wei:2008,Mao:Billings:1997,Guo:Billings:2015}. Recently, an alternate approach referred as `RJMCMC', is proposed~\cite{Baldacchino:Kadirkamanathan:2013} in which a term is either added through `\textit{birth move}' or removed through `\textit{death move}'. The probabilities for these moves are determined using the Bayesian inference. 

In contrast, in the subset search, entire candidate structure/term subset is determined and evaluated in each step. The common element in all subset search methods is the use of a stochastic component to enable better sampling of search space. Earlier research in this direction includes the application of the evolutionary algorithms such as Genetic Algorithm (GA) and Genetic Programming (GP)~\citep{Fonseca:1993,Rodriguez:Fonseca:2004,Madar:Abonyi:2005}. In recent years, various other structure selection algorithms have been proposed which also fall into the subset search category, \textit{e.g.}, see~\citep{Baldacchino:Kadirkamanathan:2012,Falsone:Piroddi:2015,Avellina:Piroddi:2017,Bianchi:Piroddi:2017}. In the randomized search proposed in~\cite{Falsone:Piroddi:2015}, known as `RaMSS', each term is assigned with a selection probability. A term is included to/excluded from the term subset on the basis of the corresponding probability. In each iteration, a fixed number of term subsets are generated and evaluated. On the basis of this evaluation, the selection probability of each term is updated. 

It is worth to mention that the goal of the structure selection is to identify a model which can capture the linear and non-linear dynamics of the underlying system while being parsimonious enough to be practical \citep{Soderstrom:Stoica:1989,Ljung:1999,Aguirre:Billings:1995,Billings:Aguirre:1995}. The major challenge for any search method is to balance these contradictory objectives, \textit{i.e., bias vs. variance trade-off}. The search for optimum structure, therefore, entails selection of \textit{correct/significant terms} as well as \textit{the number of terms}. 

This study proposes a new approach to model structure selection, where the information about the \textit{terms} and the \textit{number of terms} (referred here as \textit{cardinality}) is explicitly integrated into the search algorithm to determine the correct structure. The core idea is to use \textit{the cardinality/ the number of terms} as the additional information to balance \textit{bias-variance} dilemma. The proposed approach is, in essence, a population based search heuristic which has been developed in particle swarm theory~\citep{Kennedy:Eberhart:1995}, where each \textit{search agent (particle)} encodes a term subset/structure. In addition, each particle independently stores the \textit{selection likelihoods} of both, the \textit{cardinality (number of terms)} and the \textit{terms}. In each iteration, the best structures, obtained hitherto, are used as \textit{learning exemplars}. Further, a novel learning framework is developed to extract the information about the beneficial/significant \textit{terms} and \textit{cardinality} from the \textit{learning exemplars/best structures}. This information is used to update the \textit{selection likelihoods} of the terms and cardinality. Subsequently, a new \textit{particle/structure} is determined through the inclusive use of both the selection likelihoods. In essence, the proposed approach extends the learning dimension of the conventional particle swarms to integrate information about cardinality. Due to this distinctive property, it is referred as `Two-Dimensional (2D)' learning framework. 

Note that the 2D-learning was originally developed by the authors for the \textit{`feature selection'} problem in machine learning~\citep{Hafiz:Swain:2018}. The proof of concept of this algorithm as an alternate method of structure selection of nonlinear systems has been reported in~\citep{Hafiz:Swain:CEC:2018}. The present investigation significantly differs from~\citep{Hafiz:Swain:CEC:2018} in the following aspects: 
\begin{itemize}
    \item The proposed algorithm is better formalized and explained through illustrative example from the perspective of system identification.
    \item A detailed analysis of the search behavior and the influence of the control parameters on the search performance has been reported.
    \item The comparative evaluation is carried out considering 4-established structure selection approaches.
    \item A rigorous evaluation of the proposed approach has been carried out considering 7 benchmark nonlinear systems which are taken from the existing research.
    \item The search behavior of the algorithm, for a system with slowly varying excitation, has been investigated. This may be considered as a probable worst case scenario.
    \item A discrete time model of the continuous time system has been identified and validated through Generalized Frequency Response Functions (GFRF)~\citep{Billings:Peyton:1990}.
    \item In addition to the simulated examples, a practical case study on identification of non-linear wave-forces has been presented.
\end{itemize}

The rest of the article is organized as follows: the NARX model and the structure selection problem are briefly discussed in Section~\ref{s:pr}. The philosophy of the 2D learning is discussed in Section~\ref{s:pass}. The framework of this study is described in Section~\ref{s:IF}. The results are discussed at length in Section~\ref{s:res}, followed by the conclusions in Section~\ref{s:con}.

\section{The Structure Selection Problem}

The first step in the nonlinear system identification is the selection of system representation such as Volterra, NARX, Wiener, neural networks, polynomial models, and rational models. Among these, the NARX models yield the simplest system representation while providing precise information about system dynamics. The focus of this study is, therefore, the identification of nonlinear systems represented by polynomial nonlinear auto-regressive with exogenous inputs (NARX) models~\citep{Leontaritis:Billings:1985,LEONTARITIS:BILLINGS:1985b}. In the following, the NARX model and the structure selection problem are briefly discussed.

\subsection{The Polynomial NARX Model}
\label{s:pr}

The NARX model represents a non-linear system as a function of recursive lagged input and output terms as follows:
\begin{linenomath*}
\begin{align*}
y(k) & = F^{n_l} \ \{ \ y(k-1),\ldots,y(k-n_y),u(k-1),\ldots, u(k-n_u) \ \}+e(k)
\end{align*}
\end{linenomath*}
where $y(k)$ and $u(k)$ respectively represent the output and input at time intervals $k$, $n_y$ and $n_u$ are corresponding lags and $F^{n_l}\{ \cdotp \}$ is some nonlinear function of degree $n_l$. 

The \textit{total number of possible terms} or \textit{model size} ($N_t$) of the NARX model is given by,
\begin{linenomath*}
\begin{align}
\label{eq:Nt}
N_t & = \sum_{i=0}^{n_l} n_i, \ n_0=1 \ \textit{and \ } n_i = \frac{n_{i-1}(n_y+n_u+i-1)}{i}, \ i=1,\ldots, n_l 
\end{align}
\end{linenomath*}
This model is essentially linear-in-parameters and can be expressed as:
\begin{linenomath*}
\begin{align}
\label{eq:NARXmodel}
    y(k) & = \sum_{i=1}^{N_t} \theta_i x_i(k) + e(k)\\
    \text{where, } x_1(k) & = 1, \ \ \text{and \ } x_i(k)  = \prod_{j=1}^{p_y}y(k-n_{y_j})\prod_{k=1}^{q_u}u(k-n_{u_k}) \, \ i=2,\ldots, N_t, \nonumber
\end{align}
\end{linenomath*}
$p_y,q_u \geq 0$; $1\leq p_y+q_u \leq n_l$; $1 \leq n_{y_j}\leq n_y$;$1 \leq n_{u_k}\leq n_u$; $n_l$ is the degree of polynomial expansion.

\subsection{Problem Formulation}
Consider the identification of a nonlinear system represented by polynomial NARX model. Given a large model set with $N_t$ number of terms, denoted as,
\begin{linenomath*}
\begin{align*}
    X_{model}=\begin{bmatrix} x_1 & x_2 & \dots & x_{N_t} \end{bmatrix}
\end{align*}
\end{linenomath*}
where, $x_1, x_2, \dots x_{N_t}$ could represent any possible linear or non-linear term of the NARX model. The goal of the structure selection is to determine the \textit{correct/optimum} subset of terms, $X^{\star} \subset X_{model}$, by minimizing a suitable criterion function,`$J(\cdotp)$', \textit{i.e.},

\begin{linenomath*}
\begin{align}
    \label{eq:mss}
    J(X^{\star}) = \min \limits_{X \subset X_{model}, \ \xi_X < N_t} J(X) 
\end{align}
\end{linenomath*}
where, `$\xi_X$' denotes the number of terms present in term subset `$X$'. This will be referred as `\textit{cardinality}' in this study.

Note that, this is essentially a combinatorial optimization problem. An exhaustive search of all possible term subsets to solve this problem is often intractable even for a moderate number of terms $N_t$, as it requires the examination of $2^{N_t}$ term subsets. Since the information about cardinality (number of terms) is not known \textit{a priori}, a successful structure selection method should address the following: 1) \textit{How many terms are required to represent the system dynamics?} and 2) \textit{Which terms should be included?} These two issues are crucial to effectively address the \textit{bias-variance} dilemma.

However, most of the existing search methods focus on evaluating the significance of the terms. To the best of our knowledge, the information about the number of terms (\textit{i.e.}, \textit{cardinality}) has not been exploited for the benefit of the search process. This study, therefore, proposes a new approach with the following key features:
\begin{itemize}
    \item The information about the subset cardinality (number of terms) is determined and explicitly integrated into the search process. Instead of assigning the selection likelihoods only to the terms, these are assigned both to the \textit{number of terms (cardinality)} and the \textit{terms}. These likelihoods are updated continually throughout the search process. The new structures are explored through the joint use of these selection likelihoods.
    \item In most of the existing search approaches, a common selection probability is used to generate test structures. In contrast, in the proposed approach, each search agent/particle independently stores and updates the selection likelihoods. Thus, each particle can explore structures having different combinations of cardinality and the terms. This, in theory, often translates into the improved sampling of the search space.    
\end{itemize}
\section{Proposed Two-Dimensional (2D) Algorithm for Structure Selection}
\label{s:pass}

The Particle Swarm Optimization (PSO), which is the kernel of the 2D learning, is based on the concept of \textit{search through the social co-operation}. In essence, PSO employs a group (\textit{swarm}) of mass-less objects, referred to as `\textit{particles}', to represent a possible solution to the problem under investigation. Each particle has two memory attributes: 1) the best solution found by itself (\textit{cognitive memory}) and 2) the best solution found by the entire swarm (\textit{social memory}). These memory attributes are updated throughout the search process and used as `\textit{learning exemplars}', \textit{i.e.}, the memory attributes are used to determine a new move of each particle on the search landscape. Further details about the particle swarm optimization, its discretization framework and the 2D learning framework can be found in~\citep{Kennedy:Eberhart:1995,Hafiz:Swain:2018,Hafiz:Abdennour:2016}.

The proposed 2D learning framework has been developed based on the particle swarm theory, and this is applied to determine the correct structure of a nonlinear system represented by the polynomial NARX model. In this study, we focus on the following two key issues: 1) \textit{How many terms should be included in the structure (to be called as cardinality in the present context)}? and 2) \textit{Which terms should be included}? For this purpose, the learning dimension of particles is extended to explicitly integrate the \textit{cardinality} information into the search process, which results in the proposed 2D learning framework, as discussed in the following. 

\subsection{Philosophy of 2D Learning}
\label{s:learning}

To understand the Two-Dimensional (2D) learning in the context of system identification, consider the problem of selecting the correct structure from the NARX model with `$N_t$' number of terms. Let the set containing all model terms be denoted as $X_{model}=\begin{bmatrix} x_1 & x_2 & \dots & x_{N_t} \end{bmatrix}$. For this problem, each particle ($\beta$) encodes a \textit{candidate structure} or \textit{set of terms} in an $N_t$ - dimensional binary vector. For example, the $i^{th}$ particle, $\beta_i$, is represented as:
\begin{linenomath*}
\begin{align*}
    \beta_i & = \begin{bmatrix} \beta_{i,1} & \beta_{i,2} & \dots & \beta_{i,N_t} \end{bmatrix}, \ \ \beta_{i,m} \in \{0,1\}, \ \ m=1,2,\dots N_t
\end{align*}
\end{linenomath*}
The $m^{th}$ term ($x_m$) from $X_{model}$ is included into the candidate structure provided the corresponding bit in the particle, `$\beta_{i,m}$' is set to `$1$'. For more details, see the illustrative example in~\ref{s:appsolr}.

The key feature of the 2D learning is the explicit integration of information about the \textit{cardinality} (the \textit{number of terms}, denoted as `$\xi$') into the search process. In this learning framework, the selection likelihoods of both the \textit{cardinality} and \textit{terms} are independently stored in the `\textit{velocity}' ($V$) of each particle. For this purpose, the learning dimension of particles is extended, and the velocity of the $i^{th}$ particle is represented as a two-dimensional matrix of size $(2\times N_t)$ as follows:
\begin{linenomath*}
\begin{align}
\label{eq:vel}
V_i & =\begin{bmatrix} v_{11}^i & v_{12}^i & \dots & v_{1N_t}^i \\        
                     v_{21}^i & v_{22}^i & \dots & v_{2N_t}^i \end{bmatrix}, \ \text{where, \ } v \in \mathbb{R}
\end{align}
\end{linenomath*}
Each row of `$V$' corresponds to a separate learning dimension. The elements in the first row of $V$ store the \textit{selection likelihoods} of \textit{cardinality} (\textit{i.e.}, \textit{how many terms to be included}), and the elements in the second row give the \textit{selection likelihoods} of the corresponding terms (\textit{i.e.}, \textit{which terms to be included}). For example, `$v_{1m}^i$' gives the probability of including a total of `$m$' \textit{number of terms} in the $i^{th}$ structure, \textit{i.e.}, $\xi_i=m$. On the other hand, `$v_{2m}^i$' gives the probability of including the $m^{th}$ term in the $i^{th}$ structure. 

\begin{algorithm}[!t]
    \small
    \SetKwInOut{Input}{Input}
    \SetKwInOut{Output}{Output}
    \SetKwComment{Comment}{*/ \ \ \ }{}
    \Input{\textit{System input-output measurement data}}
    \Output{\textit{Model Structure}}
    \algorithmfootnote{The ring-topology is used to define the particle neighborhood}
    \BlankLine
    Set the search parameters: $u_f \  \& \ RG$ \\
    \BlankLine
    Randomly initialize the swarm of `$ps$' number of particles, $\{ \beta_1 \dots \beta_{ps} \}$ \\
    Initialize the velocity of each particle, $V \in \mathbb{R}_{2\times N_t}$, by uniformly distributed random numbers in [0,1] \\
    \BlankLine
    Evaluate criterion function ($J$) of each particle\\ 
    Determine the learning exemplars ($pbest$, $gbest$ and $nbest$) \\
    \BlankLine
    \For{t = 1 to iterations}
        { 
            \BlankLine
            \Comment*[h]{Swarm Update}\\
            \BlankLine
            \For{i = 1 to ps} 
            {
                \BlankLine
                \Comment*[h]{Revitalize Particles beyond Refresh Gap}\\
                \BlankLine
                \If{$ count_i \geq RG$\nllabel{line:RG1}}
                    {Re-initialize the velocity of the particle \\
                    Set $count_i$ to zero}
                \BlankLine \nllabel{line:RG2}
                Evaluate the learning sets, $\mathcal{L}_{\alpha_1}$, $\mathcal{L}_{\alpha_2}$, $\mathcal{L}_{\alpha_3}$ and $\mathcal{L}_{i}$, as per Algorithm~\ref{alg:learningset}\\
                Update the velocity of the $i^{th}$ particle as per~(\ref{eq:UPSOorig})\\
                Update the position of the $i^{th}$ particle following Algorithm-\ref{fig:posprop}
            }
            \BlankLine
            Store the old fitness of the swarm in `$\overrightarrow{J^{(t-1)}}$'\\
            \BlankLine
            Evaluate the swarm fitness, $\overrightarrow{J^t}$\\
            \BlankLine
            Update personal, global and neighborhood best positions, $pbest$, $gbest$ and $nbest$\\             
            \BlankLine
            \Comment*[h]{Stagnation Check for Refresh Gap}\\ 
            \BlankLine
            \For{i = 1 to ps\nllabel{line:RG3}} 
                { \If{$pbestval_i^{ \ t} \geq pbestval_i^{ \ t-1}$}
                        {$count_i=count_i+1$}
                } 
            \BlankLine\nllabel{line:RG4}
        }
\caption{Pseudo code of 2D-UPSO algorithm for the model structure selection}
\label{fig:2D-UPSO}
\end{algorithm}

It is worth to emphasize that the `\textit{velocity}' is the key search engine of the 2D learning. Throughout the search process, the \textit{cardinality} and term selection likelihoods, stored in the velocity, are continually updated using the information extracted from the learning exemplars. In essence, each iteration of the 2D learning involves the following three steps: 
\begin{enumerate}
    \item \textit{Evaluation of the Learning Sets}: Beneficial information about the \textit{cardinality} and the terms is extracted from the learning exemplars and stored in the learning sets.
    \item \textit{Velocity Update}: The learning sets are used to update  the \textit{cardinality} and term selection likelihoods.
    \item \textit{Position Update}: A new structure/term subset is determined through joint use of the \textit{cardinality} and term selection likelihoods.  
\end{enumerate}

The following subsections describe these procedures in detail. The overall procedure involved in 2D-UPSO is described in Algorithm~\ref{fig:2D-UPSO}. The \textit{time complexity} of this algorithm has been studied and reported in~\citep{Hafiz:Swain:2018} where it has been shown that the 2D-UPSO could complete the search in comparatively shorter \textit{run} time. 

\begin{algorithm}[!t]
    \small
    \SetKwInOut{Input}{Input}
    \SetKwInOut{Output}{Output}
    \SetKwComment{Comment}{*/ \ \ \ }{}
    \Input{Learning Exemplar ($\alpha$) and Particle Position ($\beta_i$)}
    \Output{Learning Sets: $\mathcal{L}_\alpha$ and $\mathcal{L}_i$}
    \algorithmfootnote{`$\wedge$' denotes bit-wise logical `AND' operation. `$\overline{\beta_i}$' denotes logical complement of `$\beta_i$'}
    \BlankLine
    \Comment*[h] {Learning for Subset Cardinality}\\
    Set the cardinality learning sets to an $N_t$-dimensional null vector, \textit{i.e.},
    $\varphi_{\alpha} = \varphi_{i} = \{ 0, \ 0 \dots 0 \}$ \nllabel{line:ls1}
    \BlankLine
    Determine the cardinality of the learning exemplar ($\alpha$) and the particle position ($\beta_i$): $\xi_{\alpha} = \sum \limits_{m=1}^{N_t} \alpha_{m}$ and $\xi_{i} = \sum \limits_{m=1}^{N_t} \beta_{i,m}$ \nllabel{line:ls2}\\ 
    \BlankLine
    Set the `$\xi^{th}$' bit of the cardinality learning set, `$\varphi$', to `$1$', \textit{i.e.}, $\varphi_{\alpha,\xi_{\alpha}} = 1$ and $\varphi_{i,\xi_{i}} = 1$ \nllabel{line:ls3}
    \BlankLine
    \BlankLine
    \Comment*[h] {Learning for Features}\\  
    \BlankLine
    Evaluate Term Learning Sets: $\psi_{\alpha} = \{ \alpha \wedge \overline{\beta_i} \}$ and $\psi_{i}=\beta_i$ \\ \nllabel{line:ls4}
    Evaluate the final learning sets: $\mathcal{L}_{\alpha} = \begin{bmatrix} \varphi_{\alpha} \\ \psi_{\alpha} \end{bmatrix}$ and $\mathcal{L}_{i} = \begin{bmatrix} \varphi_{i} \\ \psi_{i} \end{bmatrix}$ \nllabel{line:ls5}
    \BlankLine
\caption{Evaluation of the learning sets}
\label{alg:learningset}
\end{algorithm}
\subsection{Evaluation of Learning Sets}
\label{s:learn}
The core idea of the particle swarm theory is to \textit{learn} from the memory attributes/learning exemplars to determine better solutions. In 2D learning, this is accomplished through the extraction of learning sets from the learning exemplars. To understand this process, let `$\alpha$' denote a memory attribute/learning exemplar such as `\textit{personal best}' ($pbest$), `\textit{swarm best}' ($gbest$) or `\textit{neighborhood best}' ($nbest$). Note that, $\alpha$ is essentially the best structure (term subset) found hitherto in the search process. Thus, it is an $N_t$-dimensional binary string. 

The objective of the 2D learning is to extract the following information from the learning exemplar `$\alpha$': 1) \textit{ cardinality} and 2) \textit{the terms that have been included in `$\alpha$' but not in the particle, `$\beta$'}. Note that the information about the \textit{cardinality} and the terms is stored independently into a two-dimensional binary \textit{learning set}, $\mathcal{L}$, of size $(2\times N_t)$. A typical learning set is thus given by,
\begin{linenomath*}
\begin{align*}
\mathcal{L} & =\begin{bmatrix} \ell_{11} & \ell_{12} & \dots & \ell_{1N_t} \\                     
                     \ell_{21} & \ell_{22} & \dots & \ell_{2N_t} \end{bmatrix}, \ \text{where, \ } \ell \in \{0,1\}
\end{align*}
\end{linenomath*}
Similar to $V$, the first row of $\mathcal{L}$ stores the learning about the \textit{cardinality} and the second row stores the learning about beneficial \textit{terms} acquired from the learning exemplar.  

Given that both `$\alpha$' and `$\beta$' are essentially binary strings, the required information to determine the learning set ($\mathcal{L}$) can easily be extracted following the steps outlined in Algorithm~\ref{alg:learningset}. See~\ref{s:applearn} for an illustrative example. The similar procedure is followed to extract a learning set from each learning exemplar such `$pbest$' and `$gbest$'.

\subsection{Velocity Update}
\label{s:velupdate}
Once the learning sets have been extracted, the velocity of each particle is updated through a simple velocity update rule. For example, at any given iteration $t$, the velocity of the $i^{th}$ particle is updated as:
\begin{linenomath*}
\begin{align}
\label{eq:UPSOorig} 
V_{i}^{t+1}  & = V_{i}^t + (r_1 \times \mathcal{L}_{\alpha_1}) + (u_f r_2 \times \mathcal{L}_{\alpha_2}) + (r_3 (1-u_f) \times \mathcal{L}_{\alpha_3}) + (\Delta_{i} \times \mathcal{L}_{i})
\end{align}
\end{linenomath*}
where $u_f \in [0,1]$ denotes the \textit{unification factor}; $r_1,r_2,r_3 \in \mathbb{R}_{2 \times N_t}$ are \textit{uniformly distributed random numbers} in $[0,2]$. Further, $\mathcal{L}_{\alpha_1}$, $\mathcal{L}_{\alpha_2}$, $\mathcal{L}_{\alpha_3}$ and $\mathcal{L}_{i}$ denote respectively the learning set derived from $pbest$, $gbest$, $nbest$ and $\beta_i$.  

One of the key features of 2D learning is the inclusion of the self-learning set, $\mathcal{L}_{i}$, which is derived from the particle position, $\beta_i^t$. The influence of this learning set is controlled through `$\Delta$' which is defined as follows:
\begin{linenomath*}
\begin{align}
\label{eq:fitfeedback}
    \Delta_i & = \begin{cases}
                \delta_i, &   if \; \; J_{i}^{t}<J_{i}^{t-1}\\
                0 , &   otherwise
                 \end{cases}, \ \ \text{where, \ } \delta_i = \frac{max(\overrightarrow{J^{t}}) - J_{i}^{t}}{max(\overrightarrow{J^{t}}) - min(\overrightarrow{J^{t}})}
\end{align}
\end{linenomath*}
where, `$J_{i}^{t}$' and `$\overrightarrow{J^t}$' denote respectively the \textit{criterion function} corresponding to the $i^{th}$ particle and the entire swarm at iteration $t$. The objective here is to provide a positive feedback, `$\delta_i$', if the particle position, $\beta_i^t$, deemed beneficial. Further, the value of $\delta$ is dependent on the performance of a particle, \textit{e.g.}, the particle with the minimum fitness `$min (\overrightarrow{J^t})$' is assigned with the maximum $\delta$, as seen in~(\ref{eq:fitfeedback}).

Note that, in the proposed 2D learning framework, the velocity update rule in (\ref{eq:UPSOorig}) uses Unified Particle Swarm Optimization (UPSO)~\cite{Parsopoulos:Konstantinos:2004}, which is a well-known PSO variant. Therefore, it is referred to as `2D-UPSO' throughout this manuscript.

\subsection{Position Update}
\label{s:posupdate}
The position update procedure is carried out in two stages, to integrate the information about the cardinality and the terms, as outlined in the Algorithm~\ref{fig:posprop}. In the first stage, the cardinality of a new structure is determined. Subsequently, the required beneficial/significant terms are selected. This process is further explained through an example in~\ref{s:apppos}.
\begin{algorithm}[!t]
    \small
    \SetKwInOut{Input}{Input}
    \SetKwInOut{Output}{Output}
    \SetKwComment{Comment}{*/ \ \ \ }{}
    \Input{$v_i$; velocity of the $i^{th}$ particle}
    \Output{$\beta_i$; position of the $i^{th}$ particle }
    \BlankLine
    Initialize $\beta_i$ with an $N_t$-dimensional null vector, \textit{i.e.}, $\beta_i=\{ 0 \dots 0 \}$ \\
    \BlankLine
    \BlankLine
    \Comment*[h] {Selection of the cardinality based on roulette wheel, ($\xi_i$)}\\
    Evaluate cumulative likelihoods, $\Sigma_m = \sum \limits_{k=1}^{m} v_{1,k}^i, \ \  m = 1\dots N_t$. \nllabel{line:pos1}\\
    \BlankLine
    Determine selection probabilities, $p_m=\frac{\Sigma_m}{\Sigma_{N_t}}$
    \BlankLine
    Generate a random number, $r \in [0,1]$. \nllabel{line:pos2}\\
    \BlankLine
    Determine `$k$' such that $p_{k-1} < r <p_{k}$, this gives the cardinality of the new structure, \textit{i.e.}, $\xi_i = \{ k | p_{k-1} < r < p_k \}$ \nllabel{line:pos3}\\
    \BlankLine
    \BlankLine
    \Comment*[h]{Selection of the terms}\\
    Rank the terms on the basis of their \emph{likelihood} `$v_{2,m}^i$' and store the term rankings in vector `$\tau$' \nllabel{line:fs1}\\
    \BlankLine
    \For{m = 1 to $N_t$} 
        { \If{$\tau_{m} \leq \xi_i$}
            {$\beta_{i,m}=1$}
        } \nllabel{line:fs2}
\caption{The position update of the $i^{th}$ particle in 2D-learning}
\label{fig:posprop}
\end{algorithm}

\subsection{Criterion Function}
\label{s:CF}

As discussed earlier, each search agent in the population based meta-heuristics represents a particular term subset (\textit{e.g.}, a \textit{parent} in GA, a \textit{particle} in BPSO and 2D-UPSO). The criterion function, `$J(\cdotp)$', is essentially a performance metric which estimates the quality of the search agent/\textit{term subset} under consideration. For most of the structure selection problems, the error in model-predicted output ($\hat{y}$), over the validation data, is often considered to be a good performance metric~\citep{Stoica:Selen:2004a}. Further, it is equally important to remove the spurious terms to prevent over-fitting of the model. Hence, the structure selection problem is multi-objective in nature wherein the objective is to find a parsimonious term subset with the minimum fitting error. To balance these objectives, several information-theoretic criteria have been proposed such as Akaike Information Criterion, Bayesian Information Criterion (BIC), Minimum Description Length and the others~\citep{Stoica:Selen:2004a,Stoica:Selen:2004b}. A comparative analysis of the information criteria suggests that BIC is robust in the presence of various levels of measurement noise~\citep{Stoica:Selen:2004a}. It is therefore selected as the \textit{criterion function} in this study. 

For a given term subset, $X_i$, this is defined as:
\begin{linenomath*}
\begin{align}
\label{eq:bic}
J(X_i) & = \mathcal{N}_v \ln(\mathcal{E}) + \ln(\mathcal{N}_v) \xi_{X_i} \\
\text{where, } \mathcal{E} & = \frac{1}{\mathcal{N}_v} \sum \limits_{k=1}^{\mathcal{N}_v} [ y_k - \hat{y}_k ]^2 \nonumber
\end{align}
\end{linenomath*}
`$\mathcal{E}$' denotes the \textit{error} with respect to the \textit{model-predicted output} ($\hat{y}$) which is obtained with $X_i$; `$\mathcal{N}_v$' denotes the length of the validation data and `$\xi_{X_i}$' is the cardinality (\textit{number of terms}) in `$X_i$'. 

\begin{algorithm}[!t]
    \small
    \SetKwInOut{Input}{Input}
    \SetKwInOut{Output}{Output}
    \SetKwComment{Comment}{*/ \ \ \ }{}
    \Input{Input-output Data, $(u,y)$}
    \Output{Identified Model, Structure and Coefficients}
    \BlankLine
    \Comment*[h] {Data Pre-processing}\\
    \BlankLine
    Split the input-output data into the \textit{estimation} and \textit{validation} sets\\
    Generate set of model terms by specifying $n_u, n_y$ and $n_l$ of the NARX model \\
    \BlankLine
    \BlankLine
    \Comment*[h] {Search for the system structure}\\
    \BlankLine
    Select a meta-heuristic search algorithm (GA, BPSO or 2D-UPSO in this study)\\
    Perform $R$ independent runs of the selected search algorithm \nllabel{line:1}\\
    \BlankLine
    \For{k = 1 to $R$} 
        { \BlankLine
           Use the selected search algorithm to locate the term-subset with minimum $J(\cdotp)$\\
          \BlankLine
          Record the best term subset ($X_{run}^k$) and the corresponding criterion function ($J(X_{run}^k)$) 	found by the algorithm\\
          \BlankLine
        } 
     \BlankLine
     Select the term-subset ($X_{alg}$) with the minimum $J(\cdotp)$ out of $R$ runs, \textit{i.e.}, $X_{alg} = \{ X_{run}^m | J(X_{run}^m) = \min \limits_{k=1:R} J(X_{run}^k)$ \}\nllabel{line:2}\\
\caption{Meta-heuristic approach to the Structure Selection}
\label{fig:propss}
\end{algorithm}

\section{Investigation Framework}
\label{s:IF}

The following subsections provide the details about the framework of this investigation. In this study, the search performance of 2D-UPSO is compared with the conventional Binary Genetic Algorithm (GA), Binary Particle Swarm Optimization (BPSO)~\cite{Kennedy:Eberhart:1997}, Orthogonal Forward Regression (OFR-ERR)~\cite{Chen:Billings:1989} and OFR with pruning~\cite{Piroddi:Spinelli:2003b}. For this purpose, the search environment is set as per Section~\ref{s:ss}. The search performance of all algorithms is evaluated through $6$ test nonlinear systems which are described in Section~\ref{s:Data}. Further, the possible search outcomes are discussed in Section~\ref{s:PM}, to evaluate the term subset found by the search algorithms qualitatively. 

\subsection{Search Setup}
\label{s:ss}

The experimental setup of this investigation is described in Algorithm~\ref{fig:propss}. To accommodate the stochastic nature of GA, BPSO and 2D-UPSO, `$R$' number of independent runs of each algorithm are carried out on the test systems ($R$ \textit{is set to} $40$). Each run is set to terminate after 6000 Function Evaluations (FEs) of criterion function, $J(\cdotp)$, given by (\ref{eq:bic}). In each run, the best term subset and the corresponding $J(\cdotp)$ found by the algorithm are recorded. Due to the stochastic nature of the search algorithms, the best term subset from $R$ independent runs ($X_{alg}$) is selected, as outlined in Algorithm~\ref{fig:propss}, Line~\ref{line:1}-\ref{line:2}. 

Note that a judicious selection of search parameters is crucial to the performance of any search algorithm. The proposed 2D-UPSO has two search parameters: \textit{unification factor} ($u_f$) and \textit{refresh gap} ($RG$). The selection of these parameters and their effects on the search performance have been discussed in Section~\ref{s:resPS}. The search parameters of GA and BPSO have been selected on the basis of previous research~\citep{Kennedy:Eberhart:1997,Liu:Mei:2016} as follows: 
\begin{itemize}
    \item GA: \textit{population size = $30$; crossover and mutation probability = $[p_c,p_m]=[0.8,0.1]$}
    \item BPSO: \textit{swarm size, $ps = 30$; inertia weight, $\omega=1$; acceleration constants, $[c_1,c_2]=[2,2]$; velocity constraints, $[v_{min}, v_{max}]=[-6,6]$}
\end{itemize}
\smallskip

Further, most of the existing structure selection methods are likely to select some terms which are not present in the model, (\textit{spurious terms}). Although the number of spurious terms chosen by proposed 2D-UPSO is significantly less compared to GA and BPSO (as will be shown in Section~\ref{s:resComp}), the selection of additional terms cannot always be ruled out. A simple null-hypothesis test may be applied to the coefficients to remove such spurious terms and refine the identified structure further.

\subsection{Search Outcomes}
\label{s:PM}

For comparative evaluation purposes, the term subset found in each run of the compared algorithm is \textit{qualitatively} evaluated. Given that the proposed 2D-UPSO and the compared algorithms (GA and BPSO) are stochastic in nature, the objective of the qualitative comparison is to evaluate the robustness of the search performance. 

For this purpose, the following term sets are defined in reference to the NARX model given by (\ref{eq:NARXmodel}),
\begin{itemize}
    \item $X_{model}$ : the set containing all terms of the NARX model
    \item $X^{\star}$ : the optimum term subset or the set of system terms, $X^{\star} \subset X_{model}$
    \item $X_{run}^k$ : \textit{term subset} found in the $k^{th}$ run of the algorithm, $k = 1, 2, \dots, R$
    \item $X_{spur}^k$ : set of \textit{spurious} terms which are selected in the $k^{th}$ run of the search algorithm, but are not present in the actual system, \textit{i.e.}, $X_{spur}^k = X_{run}^k - X^{\star}$
    \item $\varnothing$ : the null set
\end{itemize}
\smallskip

On the basis of these definitions, each run of the search algorithm gives one of the following search outcomes:
\begin{enumerate}
\item \textbf{Identification of the Correct Structure} (\textit{\textbf{Exact Fitting}}) :\\
\textit{In this scenario the identified model contains all system terms and does not include any spurious terms, i.e.}, $X_{run}^k=X^{\star}$ and $X_{spur} = \varnothing$
\smallskip
\smallskip
\item \textit{\textbf{Over Fitting}} :\\
\textit{The identified model contains all system terms; however spurious terms are also selected}, \textit{i.e.}, $X_{run}^k \supset X^{\star}$ and $X_{spur} \neq \varnothing$
\smallskip
\smallskip
\item \textit{\textbf{Under Fitting-1}} :\\
\textit{The algorithm fails to identify all system terms; though it does not include any spurious terms}, \textit{i.e.}, $X_{run}^k \subset X^{\star}$ and $X_{spur} = \varnothing$
\smallskip
\smallskip
\item \textit{\textbf{Under Fitting-2}} :\\
\textit{The algorithm fails to identify all system terms; however spurious terms are selected}, \textit{i.e.}, $X_{run}^k \not \supset X^{\star}$ and $X_{spur} \neq \varnothing$
\smallskip
\end{enumerate}

It is easier to follow that, \textit{qualitatively}, the search is \textit{successful} only when the outcome is \textit{Exact Fitting} ($X_{alg}=X^{\star}$). A higher number of `\textit{exact-fitting}' search outcome indicates a robust search performance.
\subsection{Test Non-linear Systems}
\label{s:Data}

The efficacy of the search algorithms is evaluated through $6$ non-linear systems shown in Table~\ref{t:sys}. These systems have been selected from the existing investigations on the structure detection~\cite{Mendes:1995,Mao:Billings:1997,Bonin:Pirrodi:2010,Piroddi:Spinelli:2003b,Baldacchino:Kadirkamanathan:2013,Falsone:Piroddi:2015}. The systems are excited by a white noise sequence having either uniform or Gaussian distribution as shown in Table~\ref{t:sys}. A total of $1000$ input-output data points, $(u,y)$, are generated from each system and the structure selection is performed following the principle of \textit{cross-validation}; where $700$ data points are selected for the estimation purpose and the remaining data points are used for validation, \textit{i.e.}, $\mathcal{N}_v = 300$. For each system, the NARX model is generated by setting the input-output lags and the degree of non-linearity to: $[n_u,n_y,n_l]=[4,4,3]$. This gives the NARX model set, $X_{model}$, with a total of $165$ terms following~(\ref{eq:Nt}), \textit{i.e.}, $N_t=165$.

\begin{table}[!t]
  \centering
  \scriptsize
  \caption{Test Non-linear Systems}
  \label{t:sys} 
  \begin{adjustbox}{max width=0.95\textwidth}
  \begin{threeparttable}
    \begin{tabular}{ccccccccc}
    \toprule
    
    \textbf{System} & \textbf{Known Structure} & \textbf{Input}(\boldmath$u$)$^\dagger$ & \textbf{Noise} (\boldmath$e$)$^\dagger$  \\
    \midrule

    $S1$    &  $y(k) = 0.5y(k-1) + 0.3u(k-1) + 0.3u(k-1)y(k-1) + 0.5u(k-1)^2 + e(k)$     &  WUN$(0,1)$ & WGN$(0,0.002)$\\ [2ex]
    $S2$    &  $y(k) = 0.5 + 0.5y(k-1) + 0.8u(k-2) + u(k-1)^2 - 0.05y(k-2)^2 + e(k)$     &  WUN$(0,1)$ & WGN$(0,0.05)$\\[2ex]
    $S3$    &  $y(k) = 0.8y(k-1) + 0.4u(k-1) + 0.4u(k-1)^2 + 0.4u(k-1)^3 + e(k) $         &  WGN$(0,1)$ & WGN$(0,0.33^2)$\\[4ex]
    
    $S4$    &  $y(k) =$  \makecell{$ 0.1586 y(k-1) + 0.6777 u(k-1) + 0.3037 y(k-2)^2$\\[0.7ex]
                                   $ -0.2566 y(k-2) u(k-1)^2 - 0.0339 u(k-3)^3 + e(k)$} &  WUN$(0,1)$ & WGN$(0,0.002)$\\[6ex]
                                   
    $S5$    &  $y(k) = $  \makecell{$0.7 y(k-1)u(k-1) - 0.5 y(k-2)$\\[0.7ex]
    $+ 0.6 u(k-2)^2 - 0.7 y(k-2)u(k-2)^2 + e(k)$} &  WUN$(-1,1)$ & WGN$(0,0.004)$\\[6ex]
                                   
    $S6$    &  $y(k) =$  \makecell{$0.2 y(k-1)^3 + 0.7 y(k-1)u(k-1) + 0.6 u(k-2)^2$\\[0.7ex]
                    $- 0.7 y(k-2)u(k-2)^2 - 0.5 y(k-2) + e(k)$} &  WUN$(-1,1)$ & WGN$(0,0.004)$\\[6ex]
    
    \bottomrule
    \end{tabular}%
    \begin{tablenotes}
      \scriptsize
       \item $\dagger$ WUN$(a,b)$ denotes white uniform noise sequence in the interval $[a,b]$; WGN$(\mu,\sigma)$ denotes white Gaussian noise sequence with the mean `$\mu$' and the variance `$\sigma$'.  
    \end{tablenotes}
  \end{threeparttable}
  \end{adjustbox}
\end{table}%
\section{Results}
\label{s:res}

This study aims to evaluate the performance of Two-dimensional particle swarms (2D-UPSO) on the structure selection problem. For this purpose, the structure selection problem of several non-linear systems described in Section~\ref{s:Data} is considered. The selection of search parameters and its effects on the search performance of 2D-UPSO are discussed in Section~\ref{s:resPS}. For the comparative analysis purposes, four other structure selection algorithms are also considered which include: GA, BPSO, OFR-ERR and OFR-Pruning. The results of this investigation are discussed in Section~\ref{s:resComp}. 

After establishing the efficacy of 2D-UPSO, it is applied to identify the structure for two well-known continuous-time nonlinear systems, as discussed in Section~\ref{s:ResDuff}. Finally, in Section~\ref{s:WavForce}, the 2D-UPSO is applied to real-world application problem of identifying nonlinear wave force dynamics. 

\subsection{2D-UPSO: Selection of Search Parameters}
\label{s:resPS}

\begin{figure*}[!t]
\centering
\small
\begin{subfigure}{.49\textwidth}
  \centering
  \includegraphics[width=\textwidth]{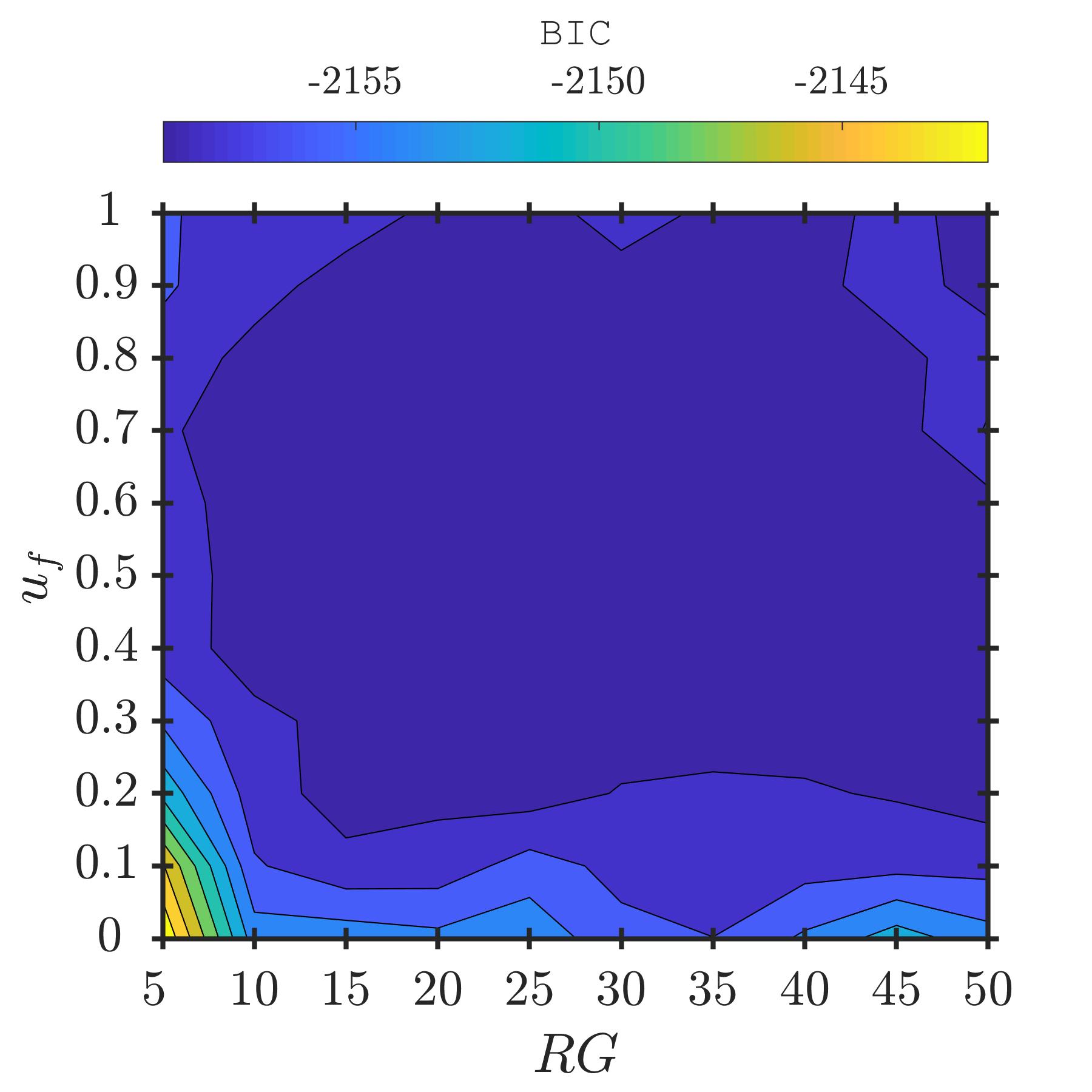}
  \caption{$S5$}
  \label{f:urg1}
\end{subfigure}%
\hfill
\begin{subfigure}{.49\textwidth}
  \centering
  \includegraphics[width=\textwidth]{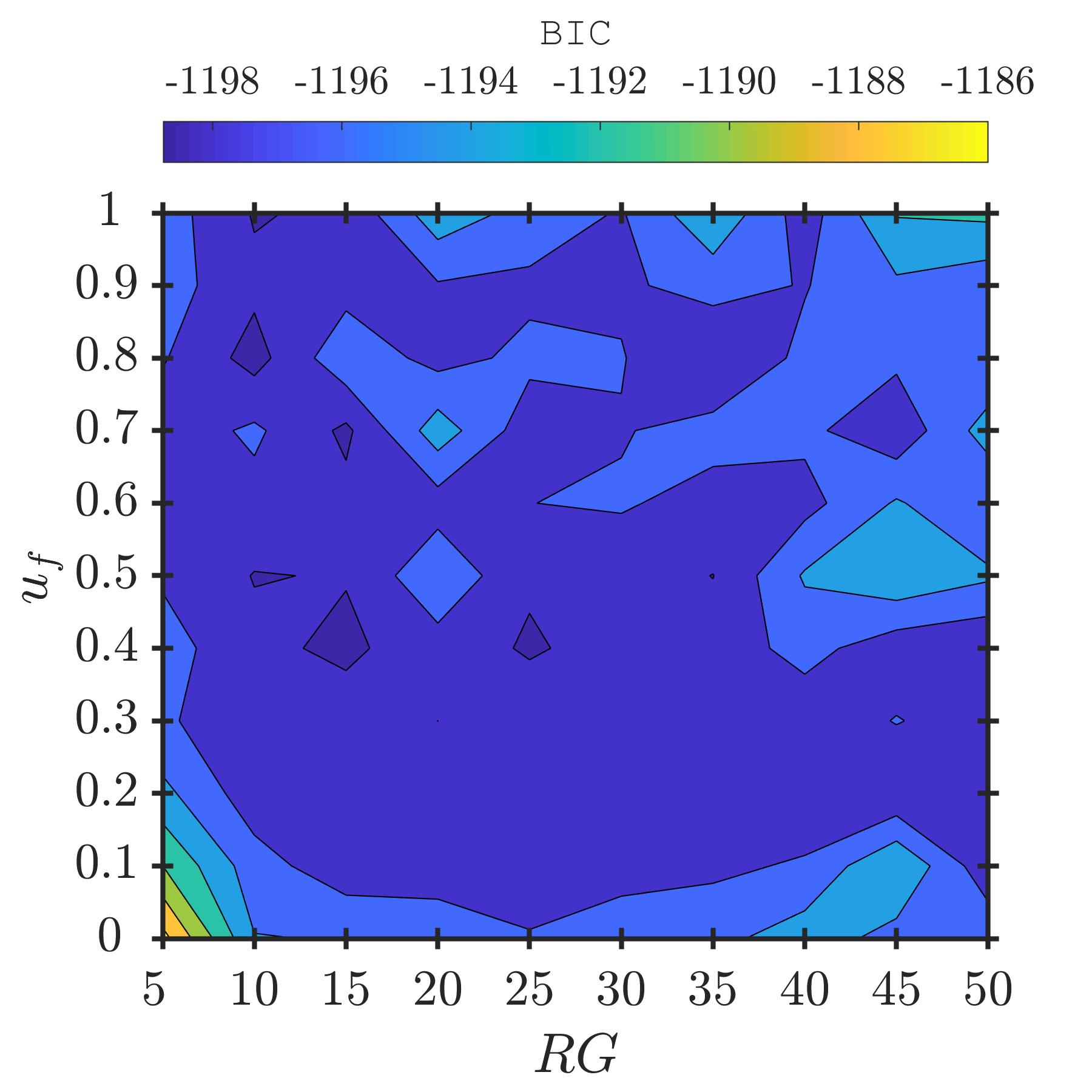}
  \caption{$S6$}
  \label{f:urg2}
\end{subfigure}

\caption{Average value of the criterion function function (BIC) over $20$ runs with different combinations of $\{u_f, RG \}$. A lower value of BIC is desirable.}
\label{f:urg}
\end{figure*}

The proposed 2D-UPSO has two tunable search parameters: the \textit{unification factor} ($u_f$) and the \textit{refresh gap} ($RG$). The unification factor essentially controls the influence of \textit{global} and \textit{local} exemplars (\textit{i.e.}, $gbest$ and $nbest$). As seen in~(\ref{eq:UPSOorig}), a higher value of $u_f$ increases the influence of the global learning set $\mathcal{L}_{\alpha_2}$ and thereby encourages \textit{exploitation} of the search space. In contrast, a lower $u_f$ encourages the \textit{exploration} of the search space as more weight is assigned to the neighborhood learning set, $\mathcal{L}_{\alpha_3}$. A detailed discussion on the neighborhood topology of swarm and its influence on the information progression and the search performance can be found in~\cite{Kennedy:Mendes:2002}.

The refresh gap, $RG$, determines the interval within which a particle can keep exploring new structures without any improvement over the best structure found hitherto by itself. The refresh gap is introduced as a preventative measure against \textit{swarm stagnation}, \textit{i.e.}, the situation in which the learning exemplars are trapped in the local minima which can eventually draw the entire swarm into the local minima. To prevent such a scenario, in 2D learning, if a particle cannot determine the better structure for a pre-fixed number of iterations (referred to as Refresh Gap, $RG$), its velocity is reinitialized, \textit{i.e.}, both \textit{cardinality} and term selection likelihoods are reinitialized to ensure that the particle can escape from the local minima. The procedure to identify the particle stagnation and velocity re-initialization is outlined in Lines~\ref{line:RG3}-\ref{line:RG4} and Lines~\ref{line:RG1}-\ref{line:RG2}, Algorithm~\ref{fig:2D-UPSO}.

To evaluate the effects of $u_f$ and $RG$ on the search performance, a grid-search has been carried out using a total of $100$ different combinations of $\{u_f, RG \}$ by varying both $u_f$ and $RG$ in the following ranges: $u_f \in [0,1]$ and $RG \in [5,50]$. For each combination of $\{u_f, RG \}$, the average value of the criterion function, $J(\cdotp)$, over $20$ independent runs, is recorded. Note that in this study the structure selection problem is designed as a minimization problem. Hence, a lower value of $J(\cdotp)$ is desirable. For the sake of brevity, only the results obtained with the system $S5$ and $S6$ are shown in Fig.~\ref{f:urg}. The results indicate that the search performance of 2D-UPSO is not very sensitive to the search parameters. Especially, relatively stable search performance is obtained for all test systems in the wide range of $u_f: [0.3-0.7]$ and $RG: [15-35]$. For the remainder of the study, these parameters are set as $u_f=0.4$ and $RG=20$.

\begin{figure*}[!t]
\centering
\small
\begin{subfigure}{\textwidth}
  \centering
  \includegraphics[width=\textwidth]{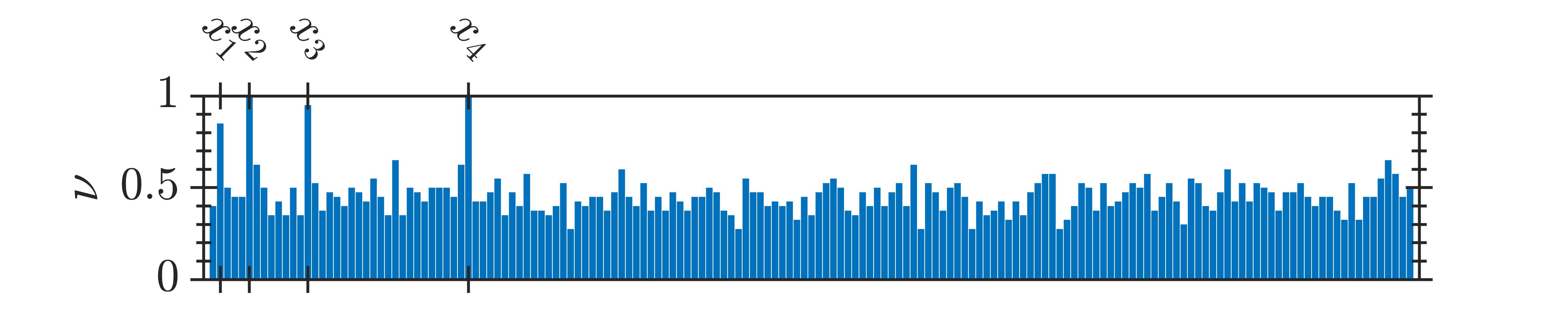}
  \caption{$S1$ + GA}
  \label{f:s1ga}
\end{subfigure}\\
\begin{subfigure}{\textwidth}
  \centering
  \includegraphics[width=\textwidth]{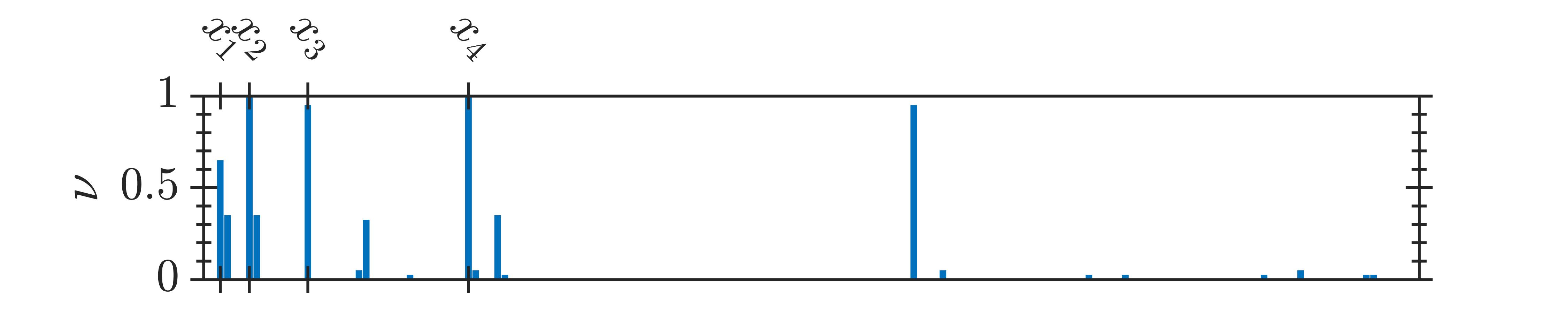}
  \caption{$S1$ + BPSO}
  \label{f:s1bpso}
\end{subfigure}\\
\begin{subfigure}{\textwidth}
  \centering
  \includegraphics[width=\textwidth]{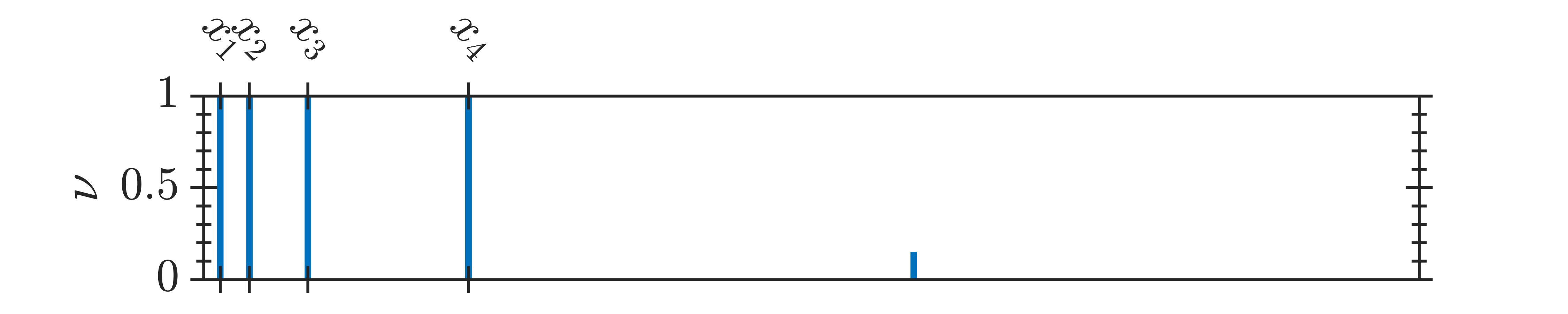}
  \caption{$S1$ + 2D-UPSO}
  \label{f:s1upso}
\end{subfigure}

\caption{Term selection frequency of system $S1$ by GA, BPSO and 2D-UPSO over 40 independent runs. The actual system terms, $\{x_1, x_2, x_3, x_4 \}$ are given in~\ref{s:app}.}
\label{f:sf1}
\end{figure*}
\subsection{Comparative Evaluation}
\label{s:resComp}

\begin{figure*}[!t]
\centering
\small
\begin{subfigure}{\textwidth}
  \centering
  \includegraphics[width=\textwidth]{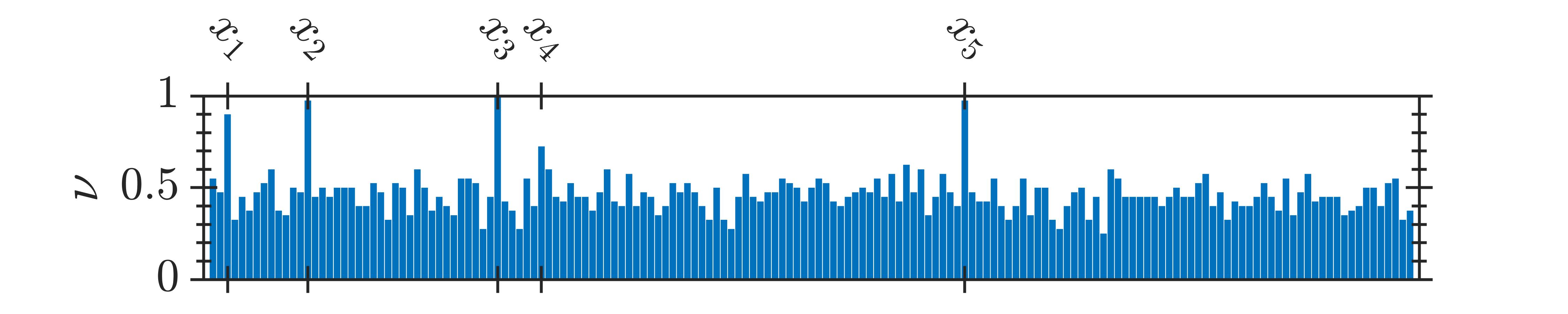}
  \caption{$S7$ + GA}
  \label{f:s7ga}
\end{subfigure}\\
\begin{subfigure}{\textwidth}
  \centering
  \includegraphics[width=\textwidth]{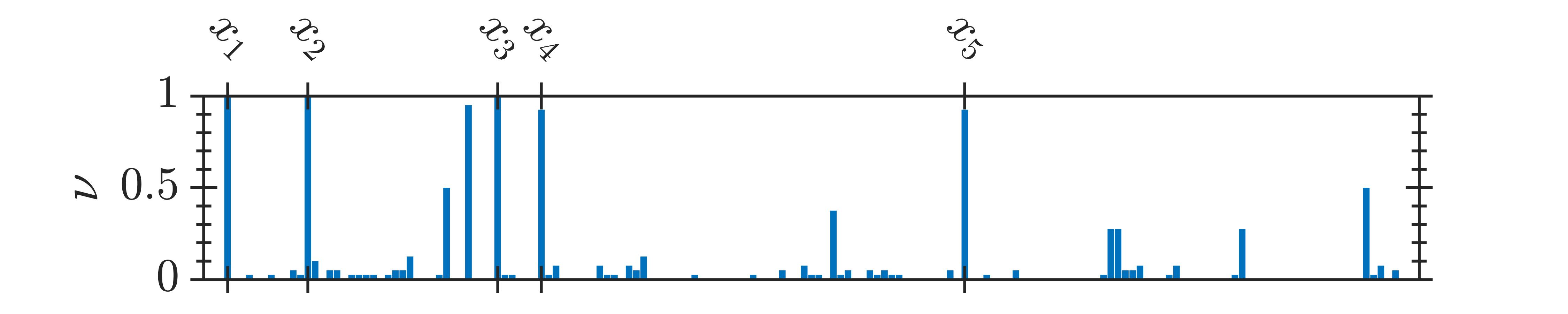}
  \caption{$S7$ + BPSO}
  \label{f:s7bpso}
\end{subfigure}\\
\begin{subfigure}{\textwidth}
  \centering
  \includegraphics[width=\textwidth]{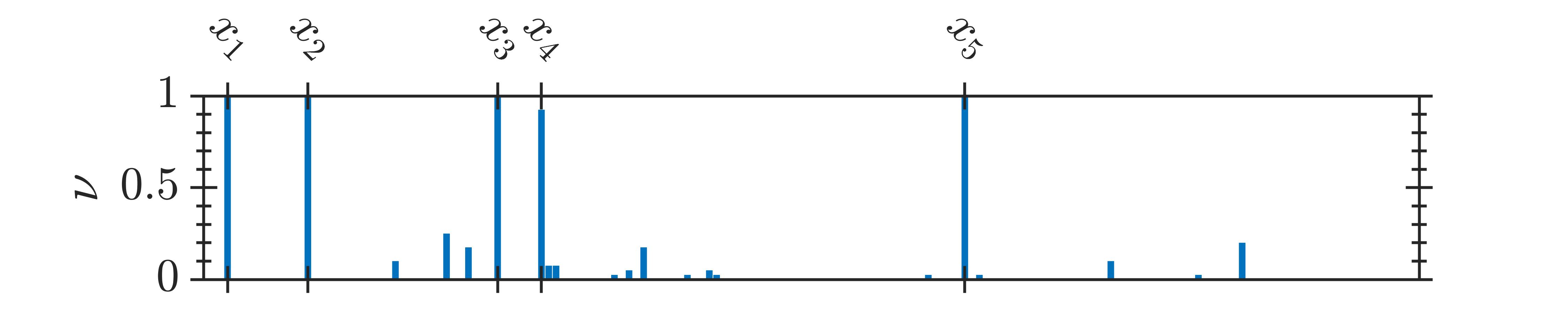}
  \caption{$S7$ + 2D-UPSO}
  \label{f:s7upso}
\end{subfigure}

\caption{Term selection frequency of system $S7$ by GA, BPSO and 2D-UPSO over 40 independent runs. The actual system terms, $\{x_1, \dots, x_5 \}$ are given in~\ref{s:app}.}
\label{f:sf2}
\end{figure*}

For each test system in Table~\ref{t:sys}, $40$ independent runs (\textit{i.e.}, $R=40$) of GA, BPSO and 2D-UPSO are carried out according to the search environment described in Section~\ref{s:ss}. For the comparative evaluation purposes, the selection frequency, `$\nu$', of each term over $R$ runs of the algorithms is determined. For the $m^{th}$ term, $x_m$, the selection frequency is evaluated as follows: 
\begin{linenomath*}
\begin{align}
\label{eq:nu}
\nu_m & = \frac{\textit{selection frequency of the } m^{th} \textit{ term}}{\textit{total number of runs}} \ = \ \frac{\Sigma_{k=1}^{R} \ a_m^k}{R} \\
\text{where, } a_m^k & = \begin{cases}
							1, \textit{if } x_m \in X_{run}^k\\
							0, \textit{otherwise}
						 \end{cases} \; \; m\in[1,N_t] \nonumber
\end{align}
\end{linenomath*}
It is easier to follow that a higher value of $\nu$ is desirable for the system terms. For all test systems, the term selection frequency with all algorithms is determined. However, for the sake of brevity, only the term selection frequency obtained for $S1$ and $S7$ is shown in Fig.~\ref{f:sf1} and~\ref{f:sf2}. Nevertheless, the search behavior for the other systems is not significantly different.

Note that at this stage, any spurious terms selected by the algorithms are not removed. Among compared algorithms, it is found that GA is prone to select a higher number of spurious terms with the selection frequency $\nu>=0.25$, as seen in Fig.~\ref{f:s1ga}-\ref{f:s7ga}. In comparison to GA, the number of spurious terms selected by BPSO as less, as seen in Fig.~\ref{f:s1bpso}-\ref{f:s7bpso}. Nevertheless, for most of the systems, both GA and BPSO failed to select some of the system terms. Further, the selection frequency of system terms with GA and BPSO is low,\textit{e.g.}, $0.52$ with $S2$ and $0.67$ with $S5$ (GA); $0.65$ with $S1$ and $0.85$ with $S5$ (BPSO). These results suggest that both GA and BPSO could yield \textit{`under-fitted'} structure (as will be discussed later in this section). 

In contrast, the selection frequency of the system terms by the proposed 2D-UPSO is significantly higher, \textit{i.e.}, $\nu>=0.97$ for all the systems. Further, 2D-UPSO could yield a significant reduction in the number, and the selection frequency of the spurious terms is obtained as seen in Fig.~\ref{f:s1upso}-\ref{f:s7upso}.

\begin{table}[!t]
  \centering
  \scriptsize
  \caption{Search Outcomes over $40$ Independent Runs$\dagger$}
  \label{t:outcome}%
  \begin{adjustbox}{max width=0.95\textwidth}
  \begin{threeparttable}
    
    \begin{tabular}{ccccc}
    \toprule
    \textbf{System} & \textbf{Outcome} & \textbf{GA} & \textbf{BPSO} & \textbf{2D-UPSO} \\
    \midrule
    \multirow{4}[2]{*}{S1} & \textit{Exact-Fitting} & 1 & 24 & \textbf{40} \\ [0.5ex]
          & \textit{Over-fitting} & 28    & 0     & 0 \\ [0.5ex]
          & \textit{Under-fitting 1} & 0     & 0     & 0 \\ [0.5ex]
          & \textit{Under-fitting 2} & $\underline{11}$    & $\underline{16}$    & 0 \\ [0.5ex]
    \midrule
    \multirow{4}[2]{*}{S2} & \textit{Exact-Fitting} & 9 & 37 & \textbf{40} \\ [0.5ex]
          & \textit{Over-fitting} & 1     & 0     & 0 \\ [0.5ex]
          & \textit{Under-fitting 1} & 2     & 0     & 0 \\ [0.5ex]
          & \textit{Under-fitting 2} & $\underline{28}$    & $\underline{3}$     & 0 \\ [0.5ex]
    \midrule
    \multirow{4}[2]{*}{S3} & \textit{Exact-Fitting} & 7 & \textbf{40} & \textbf{40} \\ [0.5ex]
          & \textit{Over-fitting} & 15    & 0     & 0 \\ [0.5ex]
          & \textit{Under-fitting 1} & 0     & 0     & 0 \\ [0.5ex]
          & \textit{Under-fitting 2} & $\underline{18}$    & 0     & 0 \\ [0.5ex]
    \midrule
    \multirow{4}[2]{*}{S4} & \textit{Exact-Fitting} & 9 & \textbf{40} & \textbf{40} \\ [0.5ex]
          & \textit{Over-fitting} & 22    & 0     & 0 \\ [0.5ex]
          & \textit{Under-fitting 1} & 0     & 0     & 0 \\ [0.5ex]
          & \textit{Under-fitting 2} & $\underline{9}$     & 0     & 0 \\ [0.5ex]
    \midrule
    \multirow{4}[2]{*}{S5} & \textit{Exact-Fitting} & 3 & 31 & \textbf{39} \\ [0.5ex]
          & \textit{Over-fitting} & 12 & 0 & 0 \\ [0.5ex]
          & \textit{Under-fitting 1} & 0 & 0 & 0 \\ [0.5ex]
          & \textit{Under-fitting 2} & $\underline{25}$ & $\underline{9}$ & $\underline{1}$ \\ [0.5ex]
    \midrule
    \multirow{4}[2]{*}{S6} & \textit{Exact-Fitting} & 4     & 30    & \textbf{37} \\ [0.5ex]
          & \textit{Over-fitting} & 23    & 4     & 0 \\ [0.5ex]
          & \textit{Under-fitting 1} & 0     & 0     & 0 \\ [0.5ex]
          & \textit{Under-fitting 2} & $\underline{13}$    & $\underline{6}$     & $\underline{3}$ \\ [0.5ex]
    \bottomrule
    \end{tabular}%

    \begin{tablenotes}
      \scriptsize
       \item $\dagger$ see Section~\ref{s:PM} for the definition of search outcomes
    \end{tablenotes}
  \end{threeparttable}
  \end{adjustbox}
\end{table}%

It is worth to emphasize that due to the stochastic nature of the algorithms, the selection of some spurious terms cannot always be ruled out. To investigate if the search performance of GA, BPSO can further be improved, the $t$-test is applied to identify and remove spurious terms from the term subset/structure obtained in each run of the algorithm, \textit{i.e.}, $X_{run}^k$. Subsequently, the refined structures are used to determine the qualitative search outcome following the definitions in Section~\ref{s:PM}. As discussed earlier, the search for the system structure can be considered successful only when all system terms have been selected, \textit{i.e.}, when the search outcome is \textit{`Exact-Fitting'} ($X_{run}^k=X^{\star}$, Section~\ref{s:PM}). 

The results in Table~\ref{t:outcome} show the number of times different categories of outcomes, such as `\textit{exact-fitting}' or `\textit{over-fitting}', are obtained over 40 runs. Note that both GA and BPSO yield under-fitted structures in many runs which indicates the absence of system terms in the selected structure. In contrast, 2D-UPSO could consistently detect true system structure in most of the runs. 


In this study, we also compared the performance 2D-UPSO with the classical structure selection approaches such as the Orthogonal Forward Regression (OFR-ERR)~\cite{Chen:Billings:1989} and OFR with pruning (OFR-Prune)~\cite{Piroddi:Spinelli:2003b} on all the systems. The limitation of the orthogonal forward regression (OFR) such as the \textit{nesting effect}, \textit{local} nature of the search, have extensively been studied in~\cite{Mendez:2001,Piroddi:Spinelli:2003,Billings:Wei:2008,Mao:Billings:1997,Guo:Billings:2015}. In contrast, the objective here is to highlight a practical limitation of OFR: the selection of proper threshold, `$\sigma$'. Both, OFR-ERR and OFR-Prune, requires a user-specified $\sigma$ which determines when the search process will terminate. Hence, the number of terms selected by OFR-ERR is critically dependent on $\sigma$. Usually, $\sigma$ is selected by trial-and-error and in conjunction with the model validity tests~\cite{Guo:Billings:2015}. 

To highlight this issue, the comparative investigation with OFR has been carried out with different values of the threshold. For the sake of brevity, only the results obtained with $S5$ and $S6$ are shown in Table~\ref{t:OFRS5} and~\ref{t:OFRS6}. Note that the search outcome is very sensitive to $\sigma$ and the determination of a proper threshold is not trivial. The improper selection of $\sigma$ can lead to either exclusion of correct system/significant term or inclusion of many spurious terms.

\begin{table*}[!t]
  \centering
  
  \begin{minipage}[c]{.49\linewidth}
  \centering 
  \scriptsize
  \caption{OFR Search Outcome ($S5^\dagger$)}
  \label{t:OFRS5}%
  \begin{adjustbox}{max width=\textwidth}
  \begin{threeparttable}
    \begin{tabular}{cccc|ccc}
    \toprule
    \multirow{3}{*}{Terms} & \multicolumn{3}{c|}{OFR-ERR} & \multicolumn{3}{c}{OFR-Prun} \\
    \cmidrule{2-7} & \multicolumn{3}{c|}{Threshold ($\sigma$)} & \multicolumn{3}{c}{Threshold ($\sigma$)} \\
    \cmidrule{2-7} & $1\%$ & $0.5\%$   & $0.12\%$ & $1\%$ & $0.5\%$ & $0.12\%$ \\
    \midrule
    $y(t-1)$         & \cmark & \cmark & \cmark & \cmark & \cmark & \cmark \\[0.8ex]
    $u(t-1)$         & \cmark & \cmark & \cmark & \cmark & \cmark & \cmark \\[0.8ex]
    $y(t-2)^2$       & \cmark & \cmark & \cmark & \cmark & \cmark & \cmark \\[0.8ex]
    $y(t-2)u(t-1)^2$ & \cmark & \cmark & \cmark & \cmark & \cmark & \cmark \\[0.8ex]
    $u(t-1)^3$       & \xmark & \xmark & \cmark & \xmark & \xmark & \cmark \\[0.8ex]
    \midrule
    $N_{spur}$       & 0 & 0 & 25 & 0 & 0 & 29 \\
    \bottomrule
    \end{tabular}%
    \begin{tablenotes}
      \scriptsize
       \item $\dagger$ `\cmark' and `\xmark' respectively denote inclusion and exclusion of system term; `$N_{spur}$' denotes total number of spurious terms
    \end{tablenotes}
  \end{threeparttable}
 \end{adjustbox}
 \end{minipage}
 \hfill
  \begin{minipage}[c]{.49\linewidth}
  \centering
  \scriptsize
  \caption{OFR Search Outcome ($S6^\dagger$)}
  \label{t:OFRS6}%
  \begin{adjustbox}{max width=0.85\textwidth}
  \begin{threeparttable}
    \begin{tabular}{cccc|ccc}
    \toprule
    \multirow{3}{*}{Terms} & \multicolumn{3}{c|}{OFR-ERR} & \multicolumn{3}{c}{OFR-Prun} \\
    \cmidrule{2-7} & \multicolumn{3}{c|}{Threshold ($\sigma$)} & \multicolumn{3}{c}{Threshold ($\sigma$)} \\
    \cmidrule{2-7} & $6\%$ & $4\%$   & $3\%$ & $6\%$ & $4\%$ & $3\%$ \\
    \midrule
   
    $y(t-2)$         & \cmark & \cmark & \cmark & \cmark & \cmark & \cmark \\ [1ex]
    $u(t-1)y(t-1)$   & \cmark & \cmark & \cmark & \xmark & \cmark & \cmark \\ [1ex]
    $u(t-2)^2$       & \cmark & \cmark & \cmark & \cmark & \cmark & \cmark \\ [1ex]
    $y(t-2)u(t-2)^2$ & \xmark & \cmark & \cmark & \xmark & \cmark & \cmark \\ [1ex]
    \midrule
    $N_{spur}$ & 1 & 1 & 26 & 0 & 0 & 46 \\

    \bottomrule
    \end{tabular}%
    \begin{tablenotes}
      \scriptsize
       \item $\dagger$ `\cmark' and `\xmark' \ respectively denote inclusion and exclusion of system term; `$N_{spur}$' denotes total number of spurious terms
    \end{tablenotes}
  \end{threeparttable}
 \end{adjustbox}
 
\end{minipage}
\end{table*}%
\subsection{Performance Under Slow Varying Input}
\label{s:slow}

In several existing investigations~\cite{Piroddi:Spinelli:2003,Guo:Billings:2015,Guo:Guo:2016,Falsone:Piroddi:2015}, a worst case scenario for identification has been used as a benchmark to show the effectiveness of their algorithms by considering a system which is excited by a slow-varying input (\textit{non-persistent excitation}). To evaluate the search performance in such a scenario, the same example has been considered here which is given by, 
\begin{small}
\begin{align}
    \label{eq:SysColor}
    \mathcal{S}_7 : w(k) & = u(k-1) + 0.5u(k-2) + 0.25u(k-1)u(k-2) -0.3u(k-1)^3\\
                    y(k) & = w(k) + \frac{1}{1-0.8z^{-1}} e(k), \ \ e(\cdotp) \sim WGN(0.02) \nonumber\\
                \text{with} \ u(k) & = \frac{0.3}{1-1.6z^{-1} + 0.64z^{-2}} v(k), \ \ v(\cdotp) \sim WGN(0,1) \nonumber
\end{align}
\end{small}
where, `$y(\cdotp)$' is the observed value of `$w(\cdotp)$'. The system is excited with a slow-varying input $u(\cdotp)$ which is essentially an AR process with a repeated pole at $0.8$. A model set consisting of 165 candidate NARX terms is generated with the following specifications: $[n_u, n_y, n_l]=[4, 4, 3]$. The system is identified by all the compared algorithms following the procedure outlined in Section~\ref{s:ss}. The search outcomes are shown in Table~\ref{t:slow}. It is observed that both BPSO and 2D-UPSO could identify all the system terms even-though only partial information is extracted from the system due to poor excitation.   

It is worth to emphasize that, this example has been considered only for the purpose of comparative analysis and such data should not be used in practice. The system should be persistently excited for identification purposes. 

\begin{table}[!t]
  \centering
  \scriptsize
  \caption{Search Outcomes for System $\mathcal{S}_7^\dagger$}
  \label{t:slow}%
  \begin{adjustbox}{max width=0.95\textwidth}
  \begin{threeparttable}
  
    \begin{tabular}{cccc}
    \toprule
    \textbf{Terms} & \textbf{GA} & \textbf{BPSO} & \textbf{2D-UPSO} \\
    \midrule
    $u(k-1)$            & \cmark    & \cmark & \cmark \\[1ex]
    $u(k-2)$            & \xmark    & \cmark & \cmark \\[1ex]
    $u(k-1)u(k-2)$      & \xmark    & \cmark & \cmark \\[1ex]
    $u(k-1)^3$          & \cmark    & \cmark & \cmark \\[1ex]
    \midrule
    $N_{spur}$          & 10     & 0     & 0 \\
    \bottomrule
    \end{tabular}%

    \begin{tablenotes}
      \scriptsize
       \item $\dagger$ `\cmark' and `\xmark' \ respectively denote inclusion and exclusion of system term; `$N_{spur}$' denotes total number of spurious terms
    \end{tablenotes}
  \end{threeparttable}
  \end{adjustbox}
\end{table}%
\subsection{Discrete Models for Nonlinear Continuous-Time Systems}
\label{s:ResDuff}

In the previous sections, several discrete polynomial NARX models have been considered to show the effectiveness of the 2D-UPSO. In this section, we apply the proposed approach to fit a discrete time model from the data obtained from continuous time systems. For this purpose, two well-known systems are considered: \textit{Duffing}'s and \textit{Van der Pol}'s Oscillator. The dynamics of these oscillators are given respectively  by,
\begin{linenomath*}
\begin{align}
	\label{eq:duffing}
	\ddot{y}(t) + 2\zeta\omega_n\dot{y}(t)+\omega_n^2y(t)+\omega_n^2\varepsilon y(t)^3-u(t)=0 \\
	\label{eq:vdp}
	\ddot{y}(t) + 2\zeta\omega_n \ (1-y(t)^2) \ \dot{y}(t)+\omega_n^2y(t)-u(t)=0
\end{align}
\end{linenomath*}
\begin{figure*}[!t]
\centering
\small
\begin{subfigure}{.4\textwidth}
  \centering
  \includegraphics[width=\textwidth]{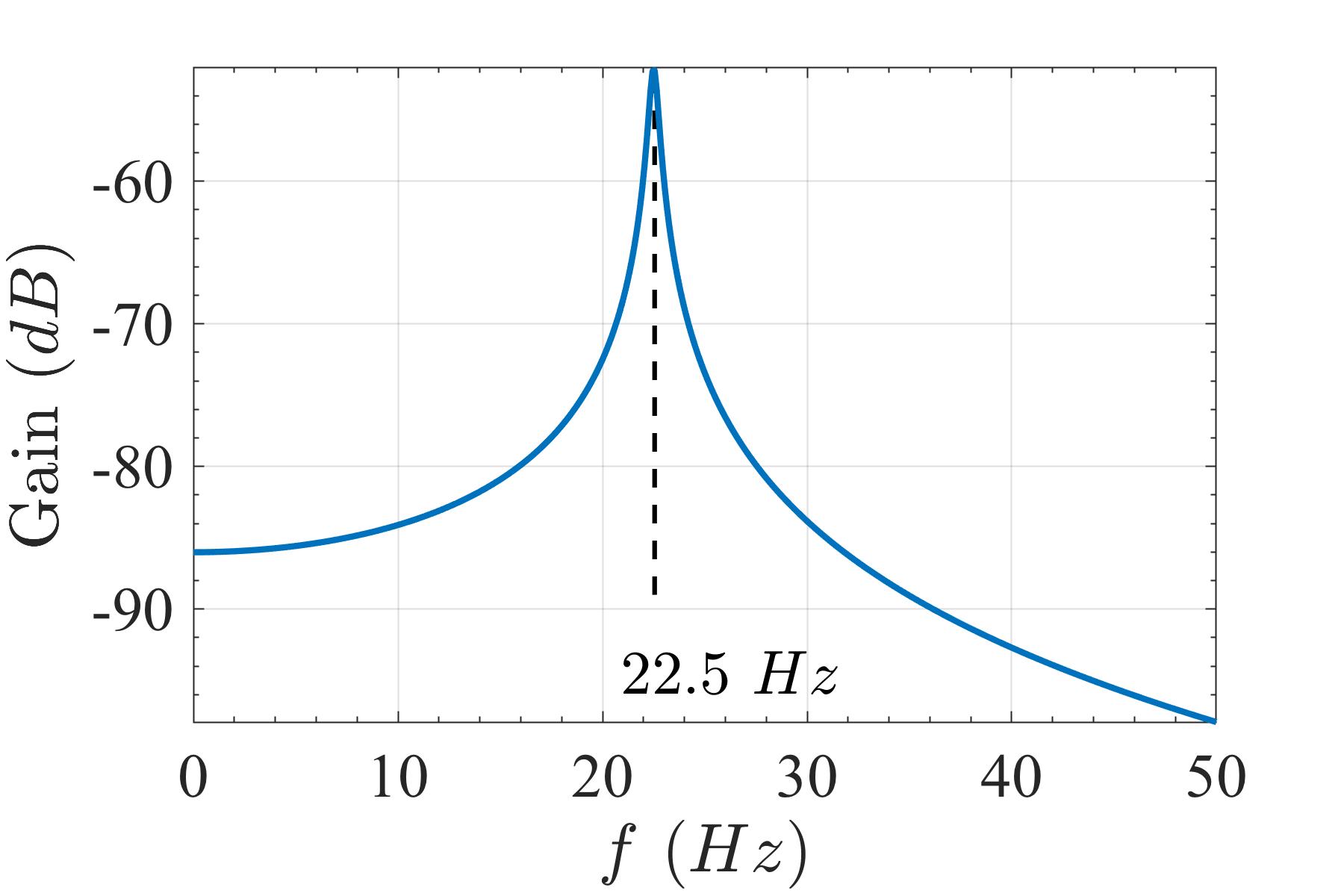}
  \caption{Linear GFRF (continuous)}
  \label{f:frfDuffC1}
\end{subfigure}%
\hfill
\begin{subfigure}{.4\textwidth}
  \centering
  \includegraphics[width=\textwidth]{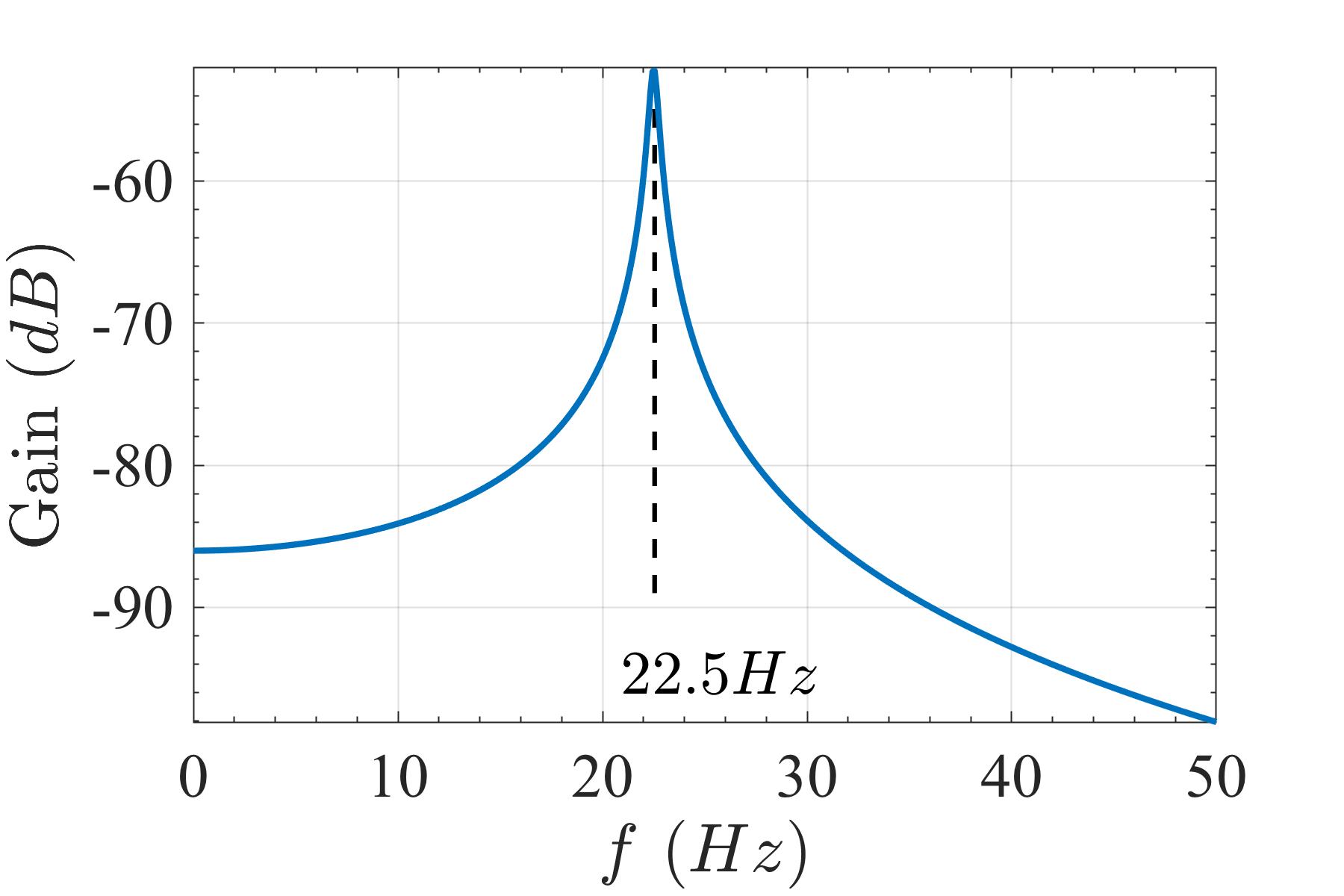}
  \caption{Linear GFRF (discrete)}
  \label{f:frfDuffD1}
\end{subfigure}
\medskip
\begin{subfigure}{.45\textwidth}
  \centering
  \includegraphics[width=\textwidth]{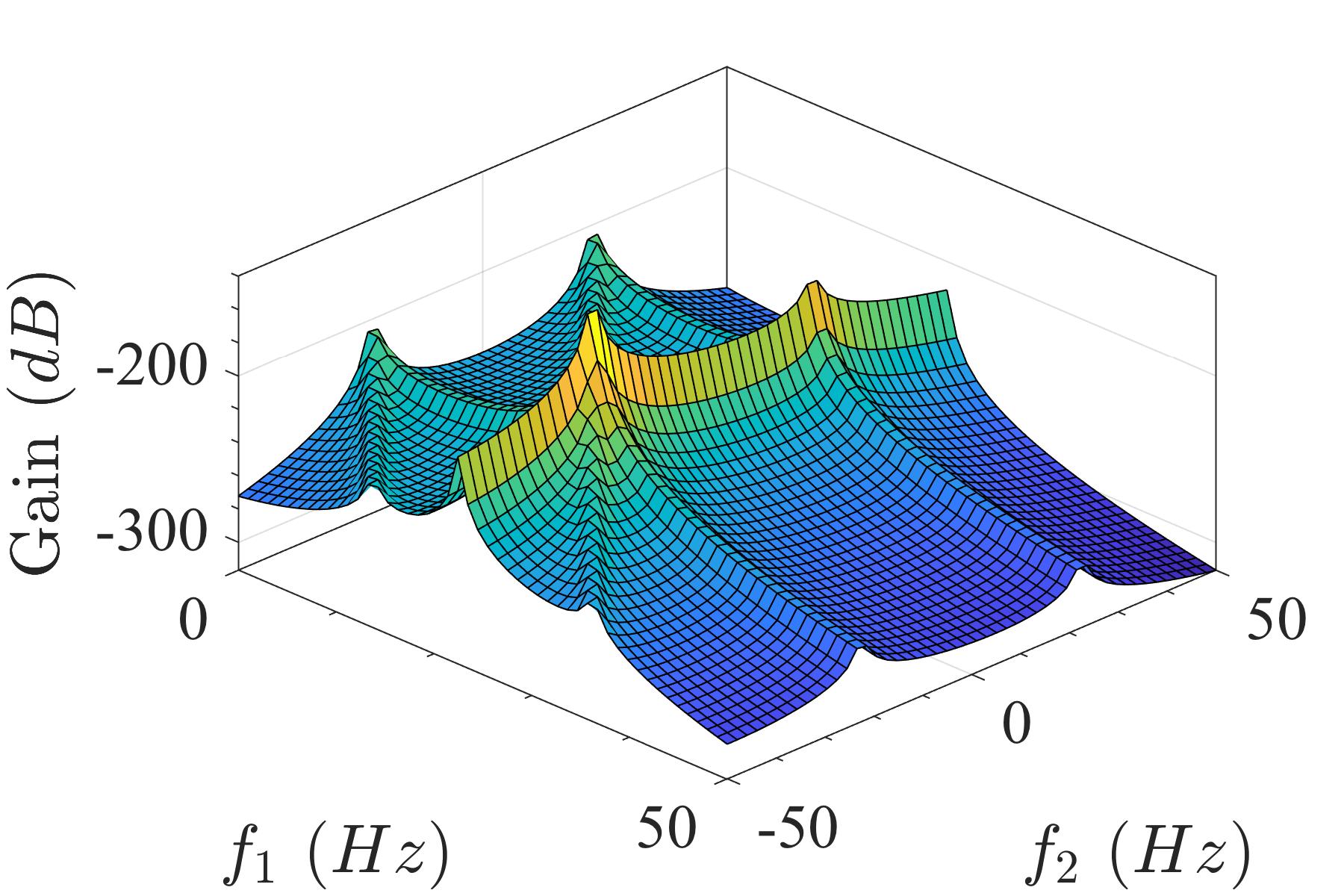}
  \caption{Third Order GFRF (continuous)}
  \label{f:frfDuffC3}
\end{subfigure}%
\hfill
\begin{subfigure}{.45\textwidth}
  \centering
  \includegraphics[width=\textwidth]{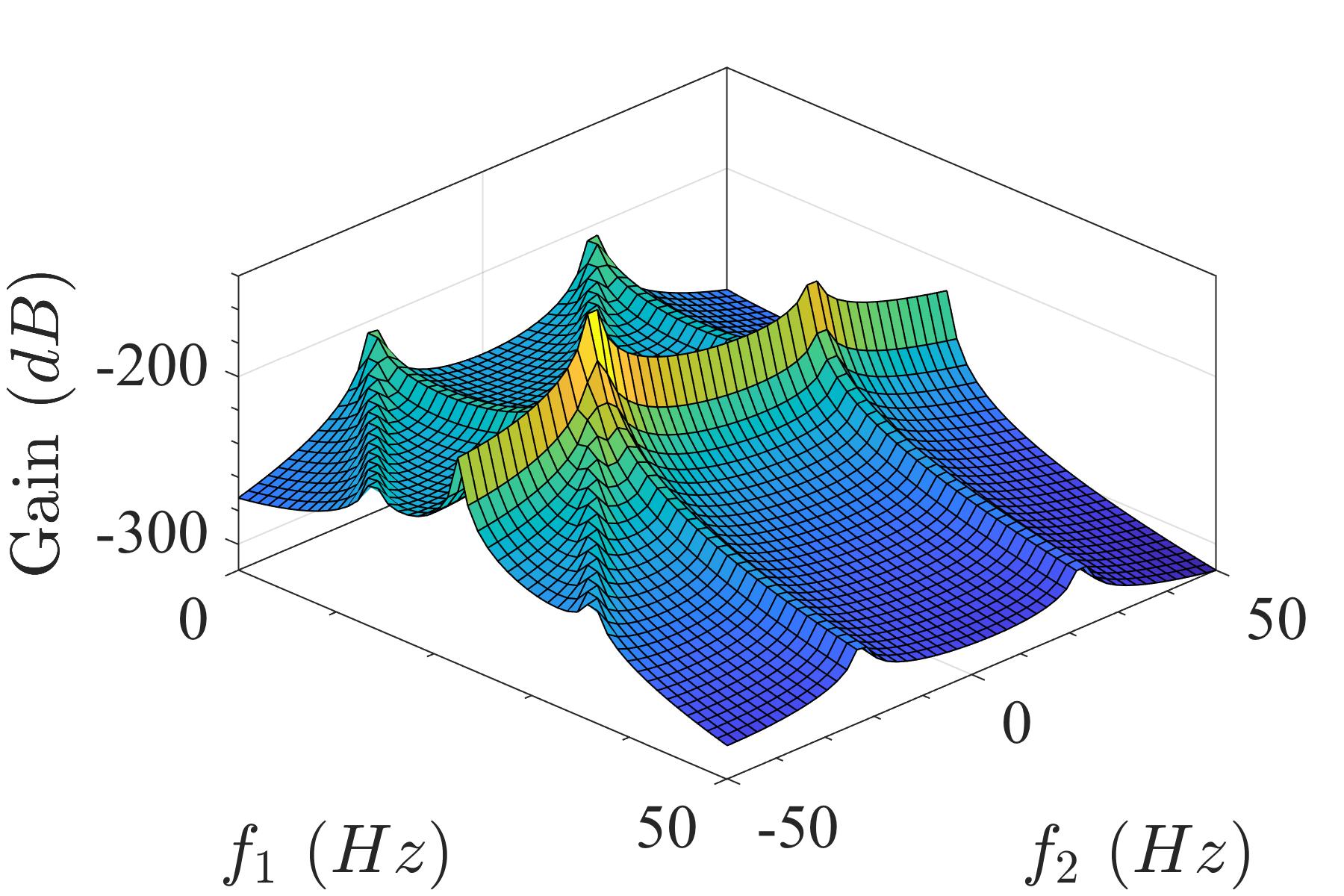}
  \caption{Third Order GFRF (discrete)}
  \label{f:frfDuffD3}
\end{subfigure}
\medskip
\begin{subfigure}{.4\textwidth}
  \centering
  \includegraphics[width=\textwidth]{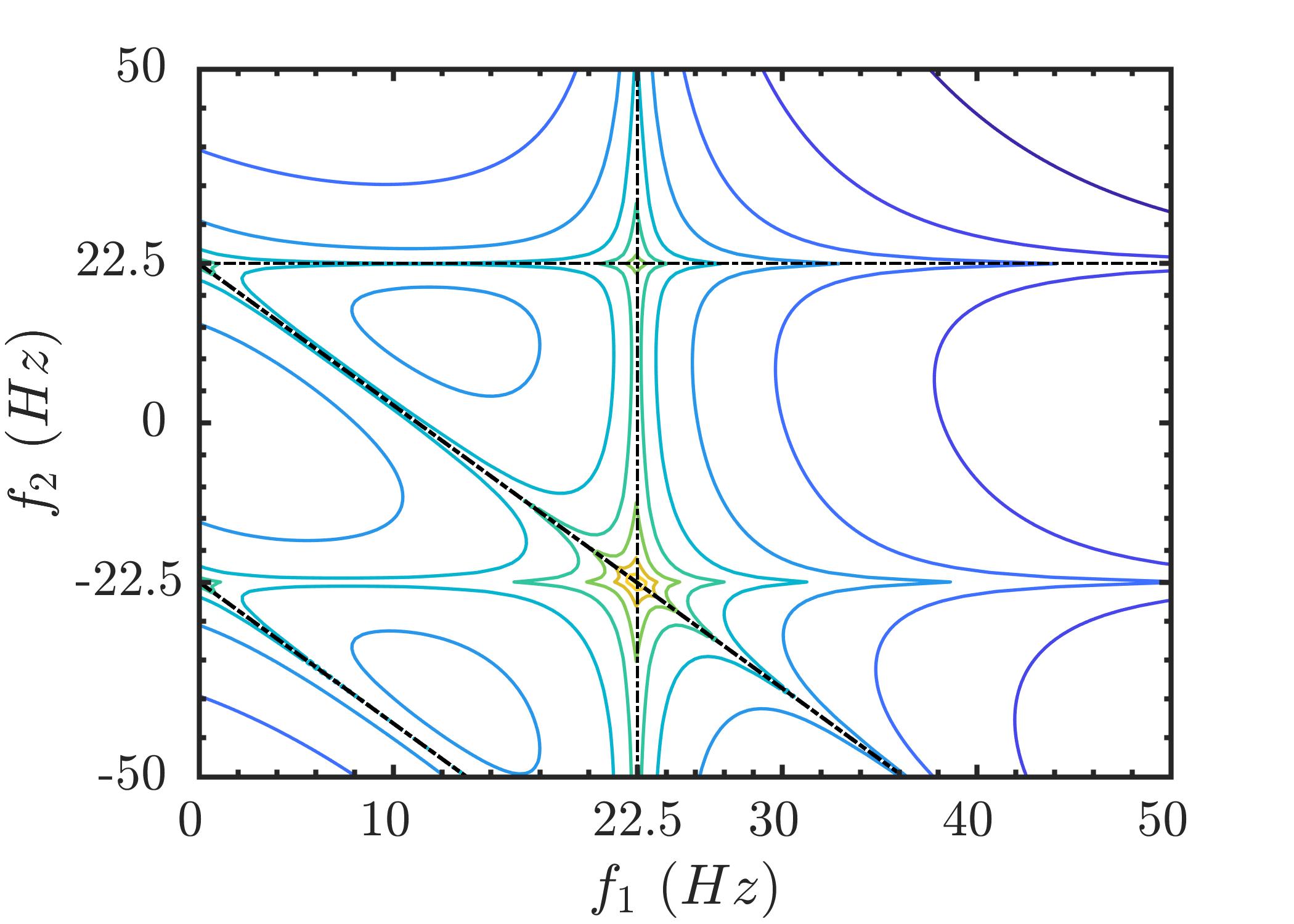}
  \caption{Third Order GFRF Contour (continuous)}
  \label{f:frfDuffC2}
\end{subfigure}%
\hfill
\begin{subfigure}{.4\textwidth}
  \centering
  \includegraphics[width=\textwidth]{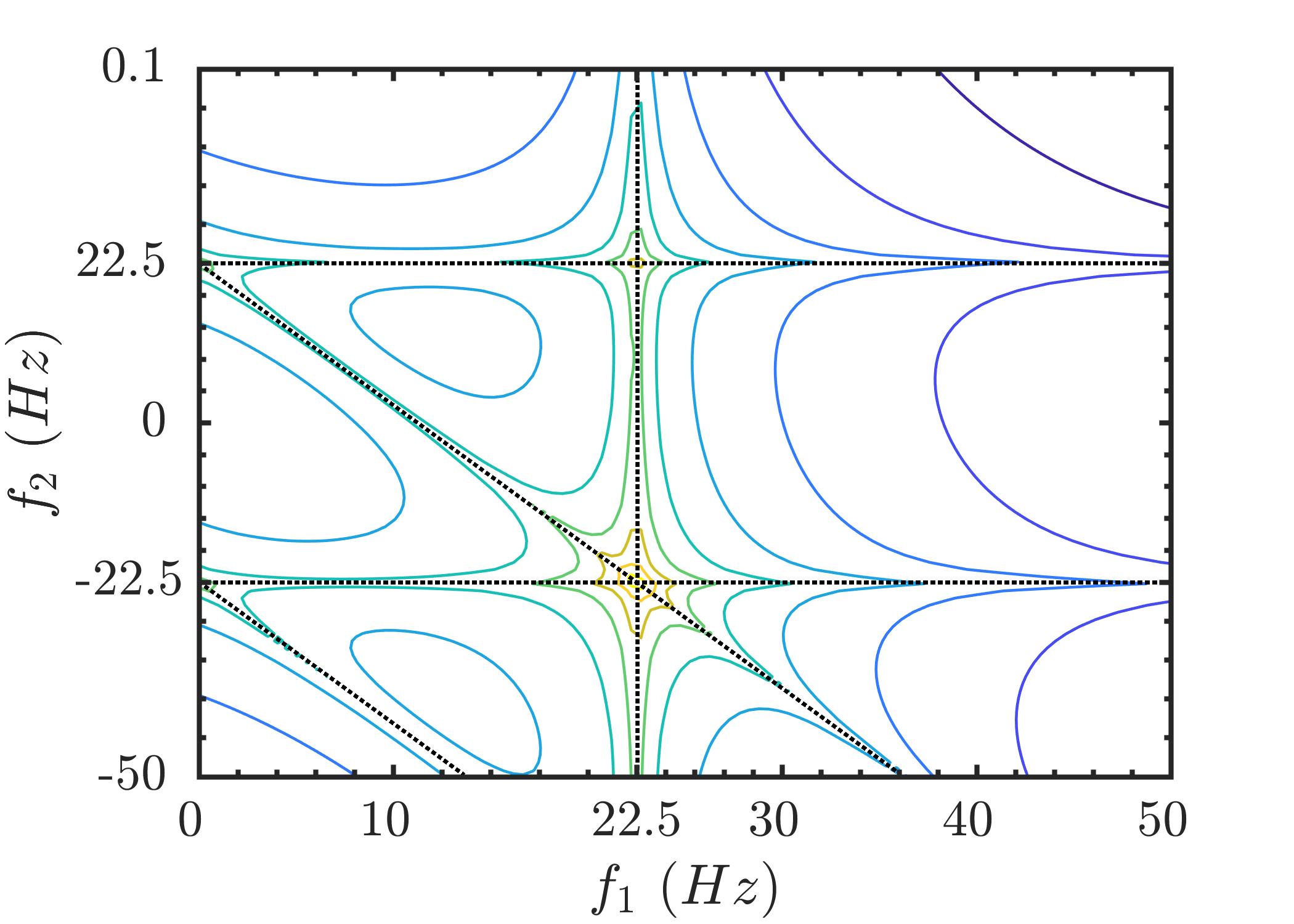}
  \caption{Third Order GFRF Contour (discrete)}
  \label{f:frfDuffD2}
\end{subfigure}
\caption{Linear and Third Order Frequency Response of Duffing's oscillator. For the third order GFRF $f_3=f_1$. Note that in the contour plots ridges align at $f_1+f_2+f_3= \pm 22.5$, which are shown by `dotted' lines in the contour plots.}
\label{f:gfrfduff}
\end{figure*}

For both models, the input-output data are generated by exciting the system with a uniformly distributed Gaussian sequence with zero mean and the output is sampled at 500 Hz. The system parameters are set as follows: $\omega_n = 45\pi$, $\zeta=0.01$ and $\varepsilon=3$. A total of $1000$ data pairs are generated. The discrete NARX model was fitted using $700$ data points, and the remaining 300 data points are used for validation.

The initial model is specified with the following NARX parameters: $n_u=5, \ n_y=5, \ n_l=3$, which gives a total of $286$ candidate terms. Following the procedure outlined in Algorithm~\ref{fig:propss}, discrete NARX models are identified for both the systems. The identified model of \textit{Duffing's} and \textit{Van der Pole's Oscillator} are shown respectively in (\ref{eq:duffIdModel}) and (\ref{eq:vdpIdModel}). 
\begin{linenomath*}
\begin{small}
\begin{align}
\label{eq:duffIdModel}                                
    y(k) & = 1.9152 \, y(k-1) - 9.9436\times 10^{-1} \, y(k-2) + 1.983\times 10^{-6} \, u(k-1)\\
         & + 1.9792\times 10^{-6} \, u(k-2) -2.5637 \times 10^{-1} \, y(k-1)^3 \nonumber\\
         & + 5.4467 \times 10^{-2}  \, y(k-3)y(k-1)^2 - 3.191 \times 10^{-2} \, y(k-1)y(k-3)^2 \nonumber\\
         \medskip
\label{eq:vdpIdModel} 
    y(k) \ = & \  1.6792 \, y(k-2) - 9.8875 \times 10^{-1} \, y(k-4) + 1.983\times 10^{-6} \, u(k-1)\\ \nonumber
            & + 5.777\times 10^{-6} \, u(k-2) + 5.7624\times 10^{-6} \, u(k-3) + 1.9681\times 10^{-6} \, u(k-4) \\\nonumber            
            & + 3.9477 \times 10^{-3} \, y(k-1)^3 + 2.7116 \times 10^{-3} \, y(k-5)y(k-3)y(k-2)\\\nonumber
            & - 2.9691 \times 10^{-4} \, y(k-5)y(k-1)^2 - 3.0699 \times 10^{-3} \, y(k-3)y(k-2)^2\\ \nonumber
            & - 3.4189 \times 10^{-3} \, y(k-5)y(k-3)^2 \nonumber
\end{align}  
\end{small}
\end{linenomath*}
The \textit{Normalized Mean-Squared Error} (NMSE) in model predicted output ($\hat{y}$) of the identified models is: $3.96\times10^{-6}$ (for \textit{Duffing's Oscillator}) and $5.87\times10^{-9}$ (for \textit{Van der Pol's Oscillator}). 


\subsubsection{Model Validation using Generalized Frequency Response Functions}
\label{s:ResGFRF}

\begin{figure*}[!t]
\centering
\small
\begin{subfigure}{.4\textwidth}
  \centering
  \includegraphics[width=\textwidth]{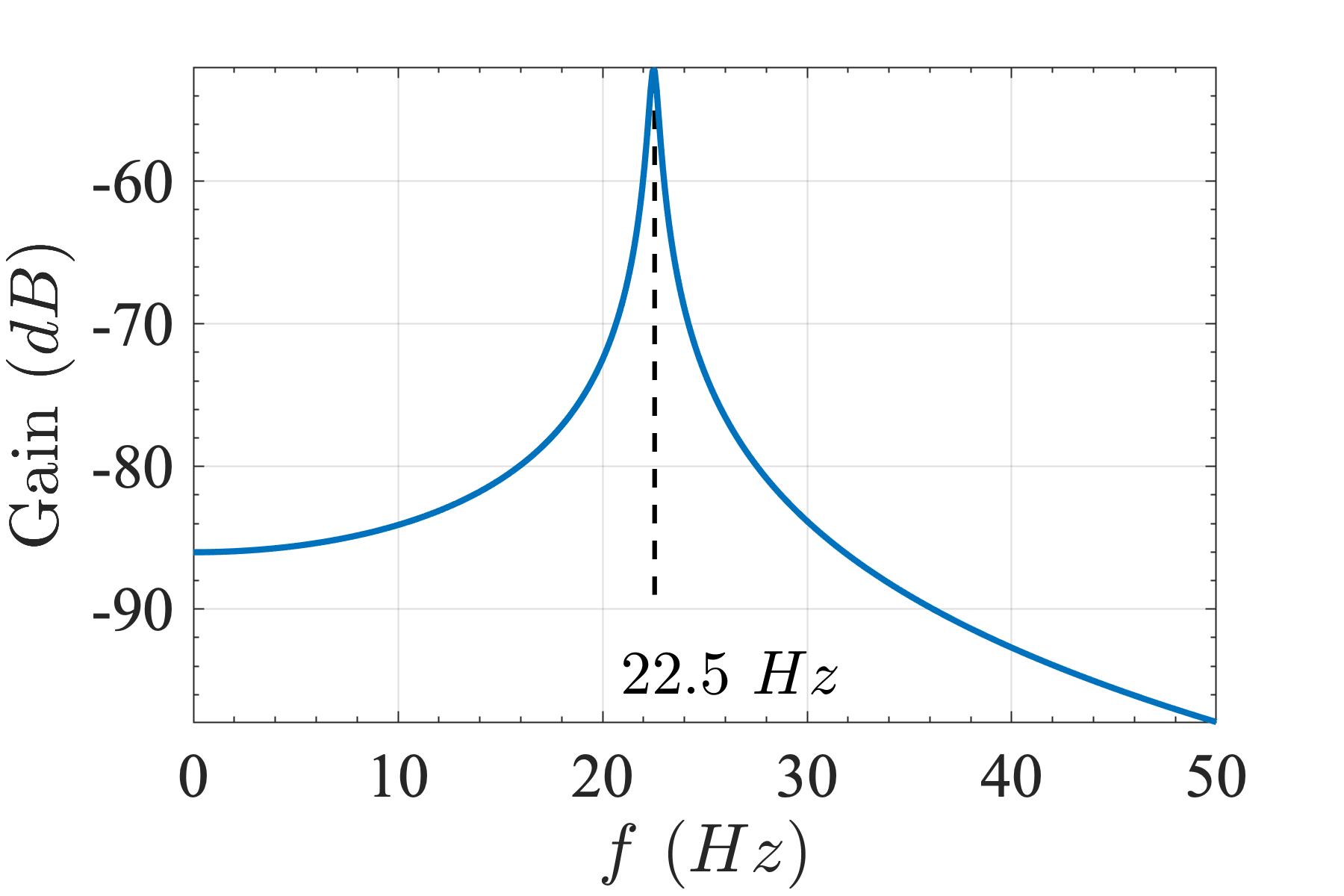}
  \caption{Linear GFRF (continuous)}
  \label{f:frfVdpC1}
\end{subfigure}%
\hfill
\begin{subfigure}{.4\textwidth}
  \centering
  \includegraphics[width=\textwidth]{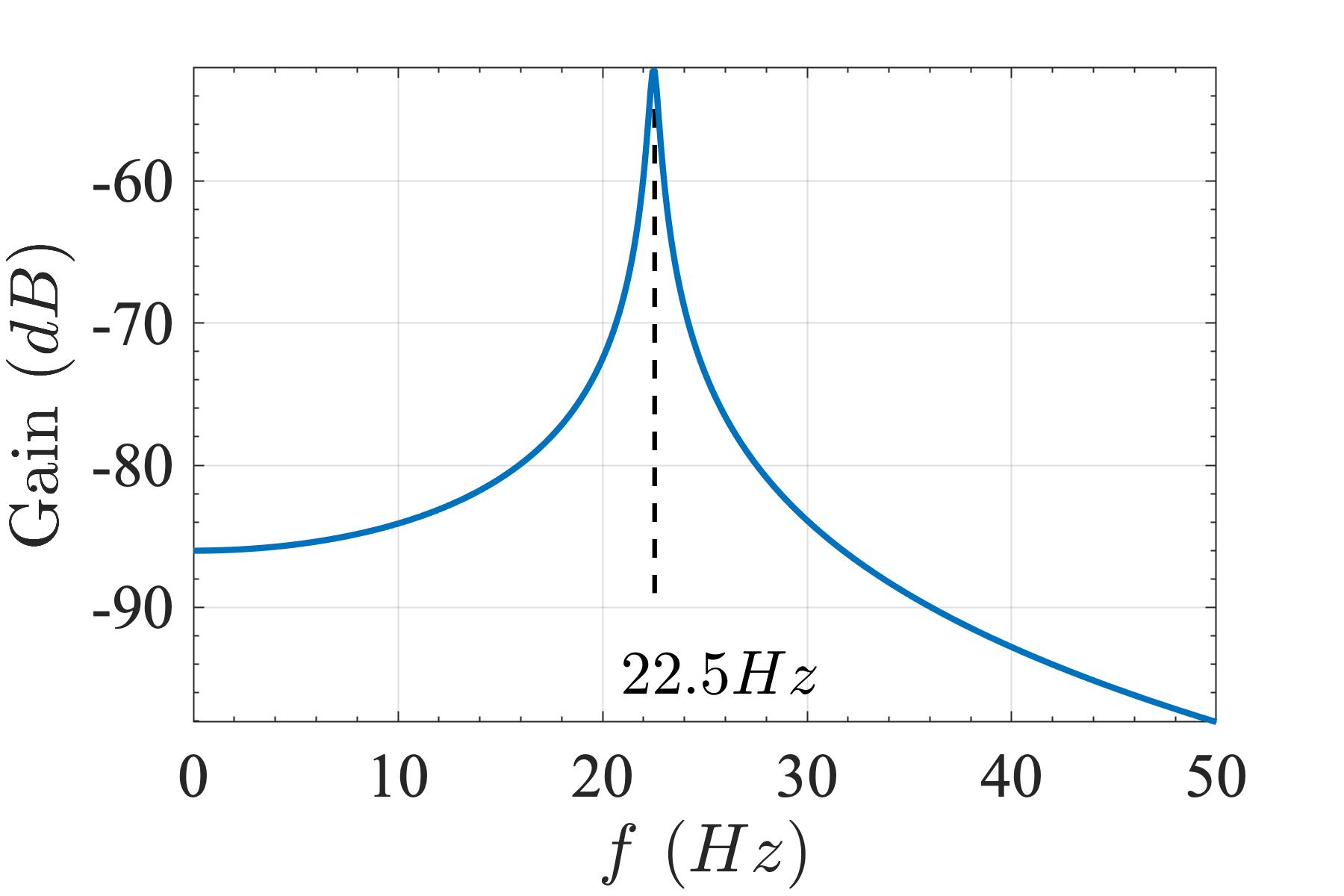}
  \caption{Linear GFRF (discrete)}
  \label{f:frfVdpD1}
\end{subfigure}
\medskip
\begin{subfigure}{.45\textwidth}
  \centering
  \includegraphics[width=\textwidth]{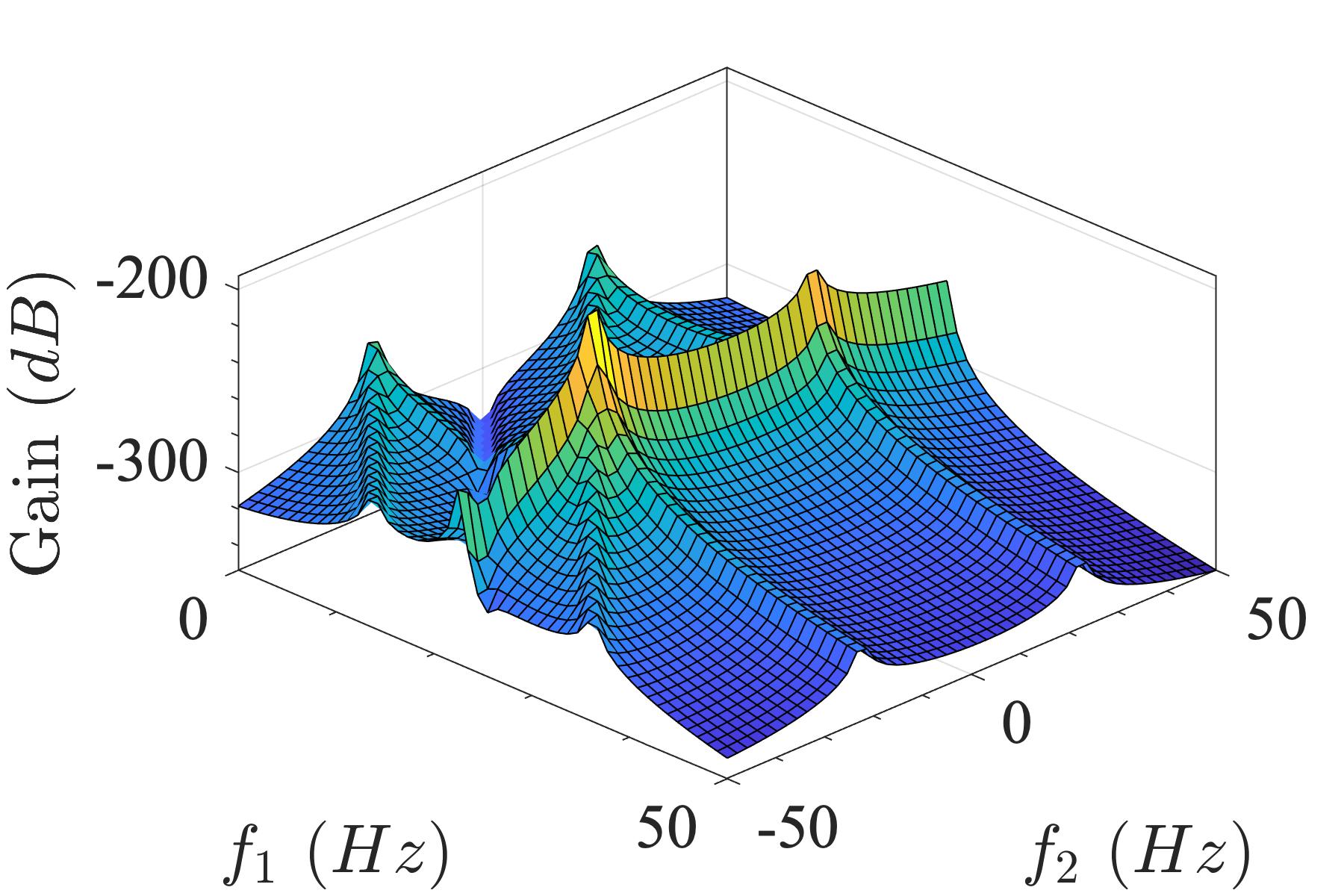}
  \caption{$3^d$ Order GFRF (continuous)}
  \label{f:frfVdpC3}
\end{subfigure}%
\hfill
\begin{subfigure}{.45\textwidth}
  \centering
  \includegraphics[width=\textwidth]{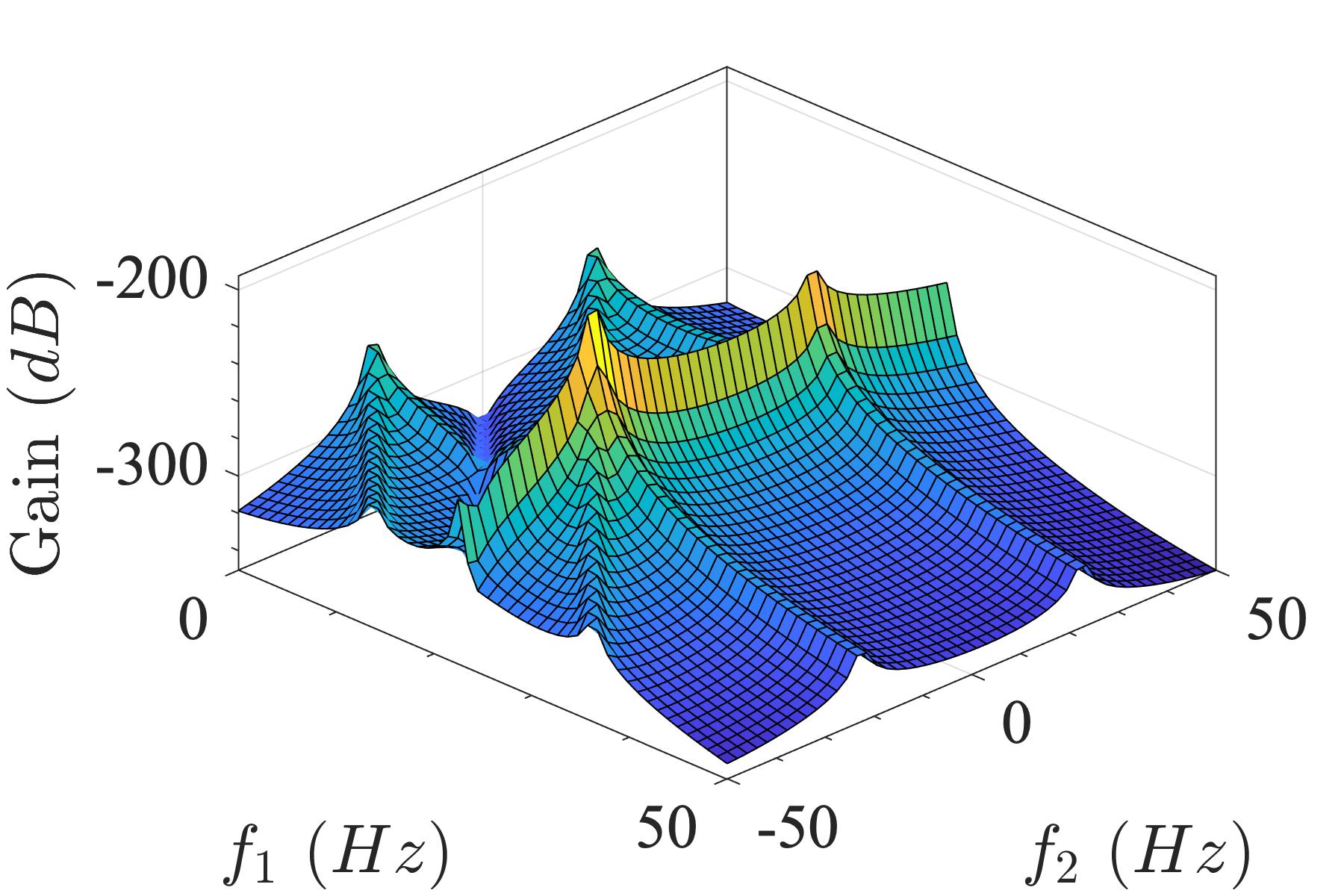}
  \caption{$3^d$ Order GFRF (discrete)}
  \label{f:frfVdpD3}
\end{subfigure}
\medskip
\begin{subfigure}{.4\textwidth}
  \centering
  \includegraphics[width=\textwidth]{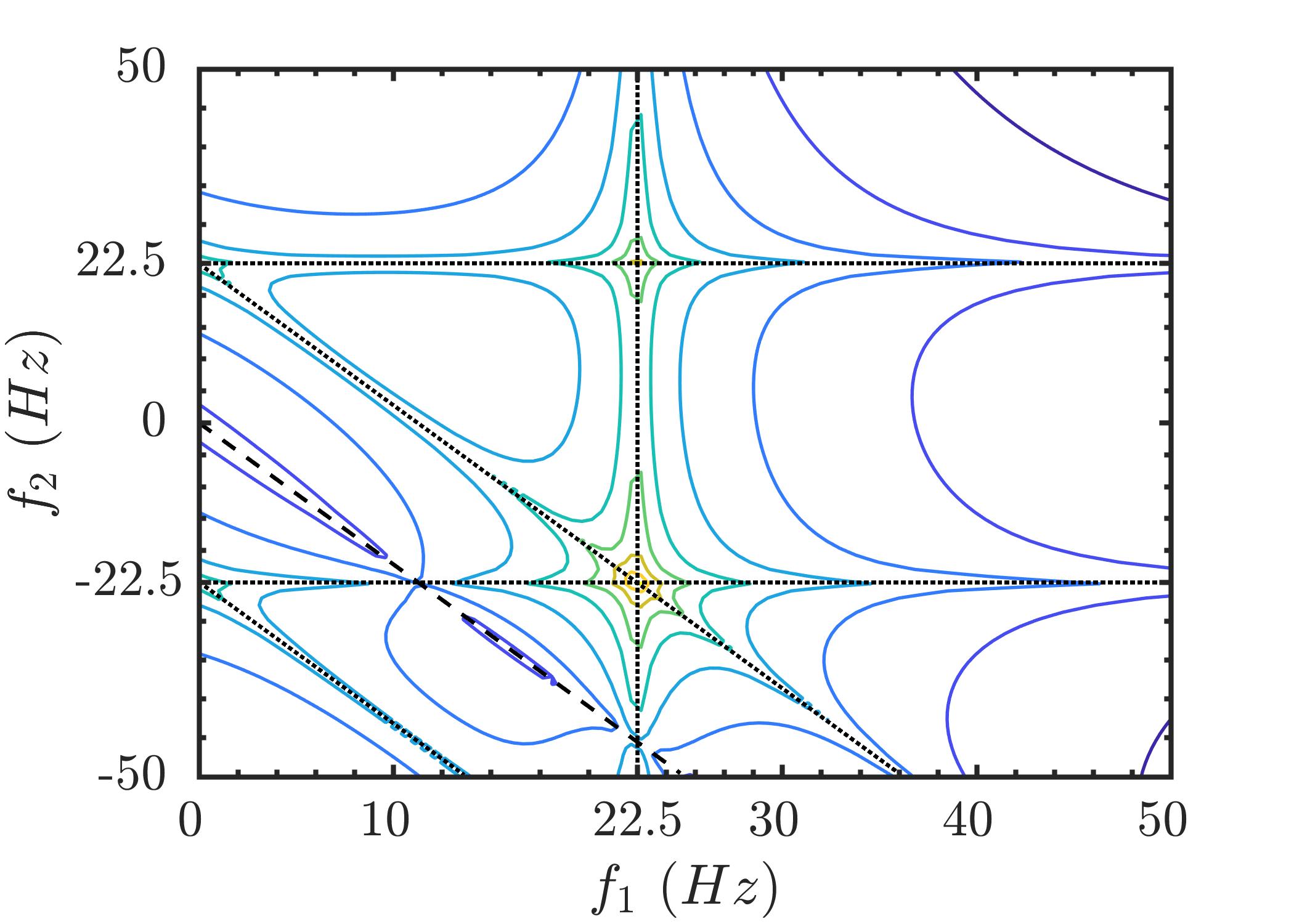}
  \caption{$3^d$ Order GFRF Contour (continuous)}
  \label{f:frfVdpC2}
\end{subfigure}%
\hfill
\begin{subfigure}{.4\textwidth}
  \centering
  \includegraphics[width=\textwidth]{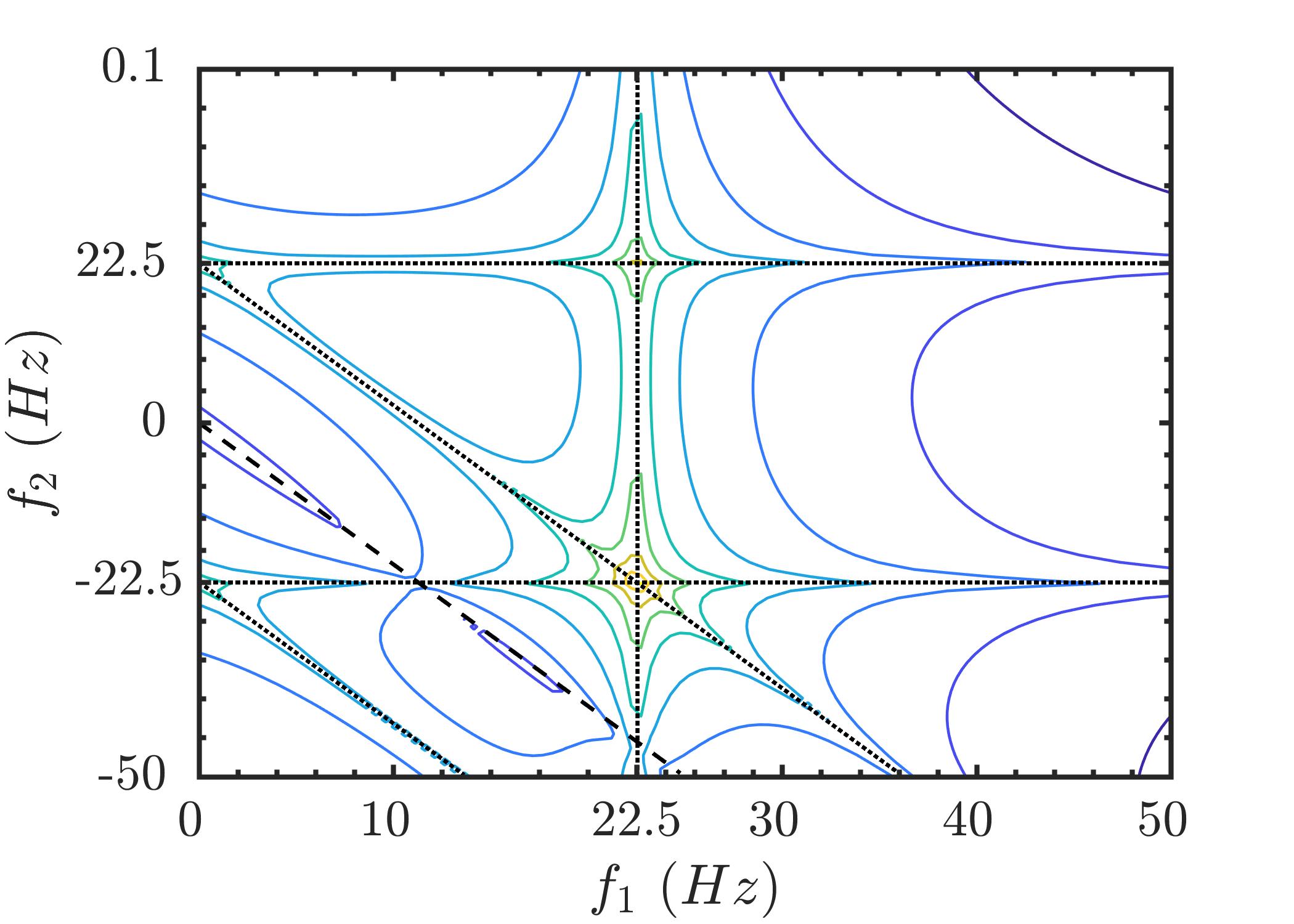}
  \caption{$3^d$ Order GFRF Contour (discrete)}
  \label{f:frfVdpD2}
\end{subfigure}
\caption{Linear and Third Order Frequency Response of Van der Pol's oscillator. For the third order GFRF $f_3=f_1$. The peaks and ridges align at $f_1+f_2+f_3= \pm 22.5$, which are shown by `dotted' lines in the contour plots. Also see the `gorge' shown by `dashed' line in the contour plots. This a typical feature of the Van der Pol and aligns at  $f_1+f_2+f_3= 0$.}
\label{f:gfrfvdp}
\end{figure*}

The validation of the model is an essential step in the system identification. In all the examples considered in the previous sections the correct structure of the system being investigated was known. Therefore it was straightforward to see whether the correct terms are included or not. However, this is not the case when the data is generated from a continuous time system.

The model is therefore initially validated both by comparing the model predicted output and cross-correlation based model validity tests~\citep{Billings:Voon:1986}. In this study, we further establish the accuracy of the models by comparing the Generalized Frequency Response Functions (GFRFs). Note that the discrete model of a continuous time system is not necessarily unique. However, if the discrete model has correctly captured the dynamics of the underlying continuous-time system, then the frequency response functions corresponding to the discrete model should match with those of the continuous time system. This, therefore, has been the motivation to show the efficacy of the proposed approach by comparing GFRFs of the discrete model with the continuous system.

The linear and the third order GFRFs of the continuous time models are shown in Fig.~\ref{f:frfDuffC1},~\ref{f:frfDuffC3} and ~\ref{f:frfDuffC2} (for Duffing's) and in Fig.~\ref{f:frfVdpC1},~\ref{f:frfVdpC3} and \ref{f:frfVdpC2} (for Van der Pole's). For the third order GFRF $f_3=f_1$. Note that the GFRF of both the oscillators is characterized by several peaks and ridges as seen in Fig.~\ref{f:frfDuffC3},~\ref{f:frfDuffC2},~\ref{f:frfVdpC3} and \ref{f:frfVdpC2}. The ridges occur both at the linear resonant frequency ($\pm 22.5 \ Hz$ for both oscillators) and when $f_1+f_2+f_3= \ \pm 22.5$. The peaks occur when all $f_1$, $f_2$ and $f_3$ are equal to  $\pm 22.5 \ Hz$. Further distinguishing characteristic of Van der Pole's oscillator is the presence of `gorge' along the line given by $f_1+f_2+f_3= \ 0$, as seen in Fig.~\ref{f:frfVdpC3} and \ref{f:frfVdpC2}. It is worth to note that these results are in agreement with the earlier investigation carried out by Billings and Jones~\cite{Billings:Peyton:1990}.

The frequency response of the discrete identified models are shown in Fig.~\ref{f:frfDuffD1},~\ref{f:frfDuffD3} and ~\ref{f:frfDuffD2} (for Duffing's) and in Fig.~\ref{f:frfVdpD1},~\ref{f:frfVdpD3} and \ref{f:frfVdpD2} (for Van der Pole's). It is clear that both the identified models could accurately capture the frequency response of the continuous models. 

\subsection{Identification of Nonlinear Wave Forces}
\label{s:WavForce}

\begin{figure*}[!t]
\centering
\small
\begin{subfigure}{.5\textwidth}
  \centering
  \includegraphics[width=\textwidth]{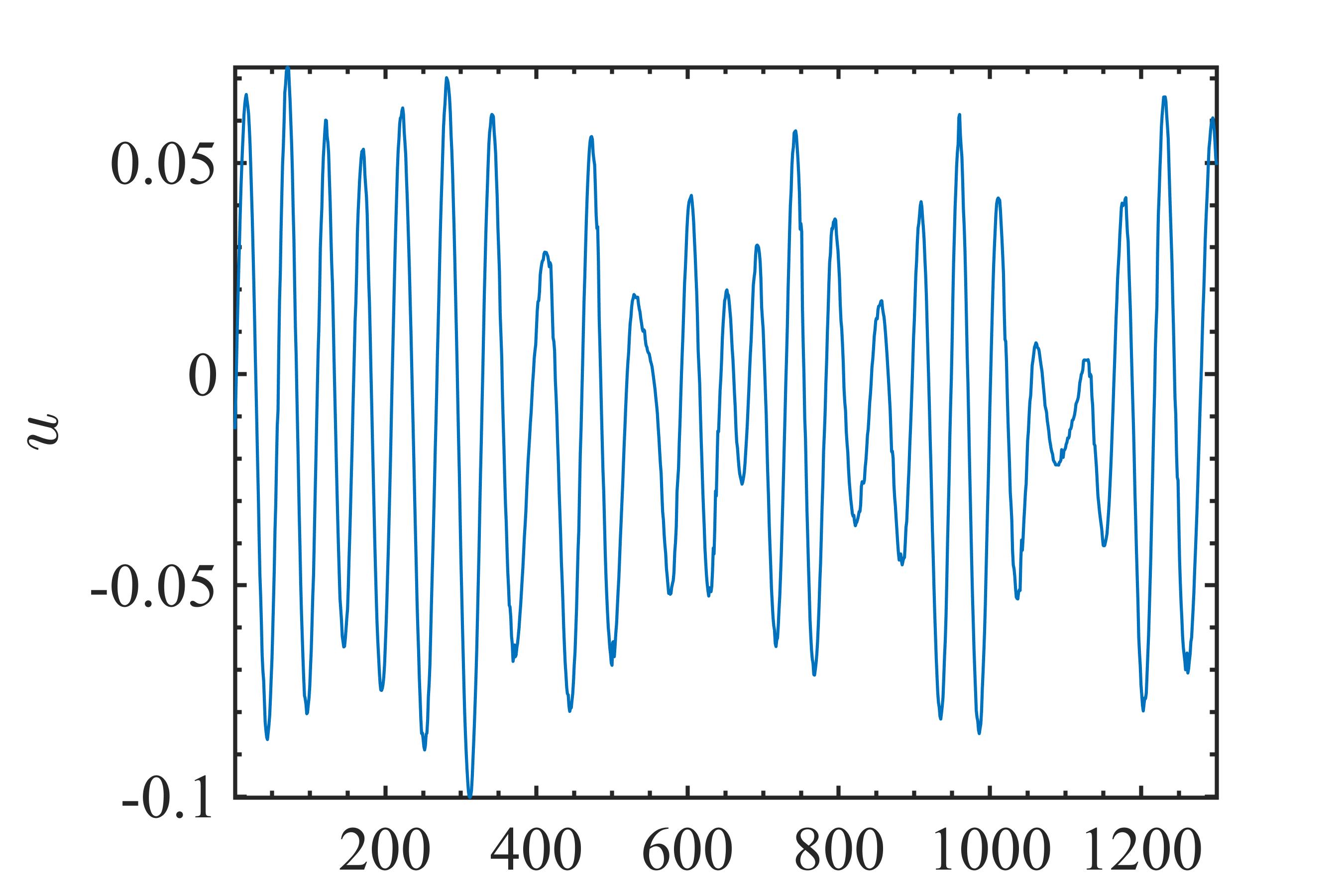}
  \caption{In-line force (input), $u$}
  \label{f:uwav}
\end{subfigure}%
\hfill
\begin{subfigure}{.5\textwidth}
  \centering
  \includegraphics[width=\textwidth]{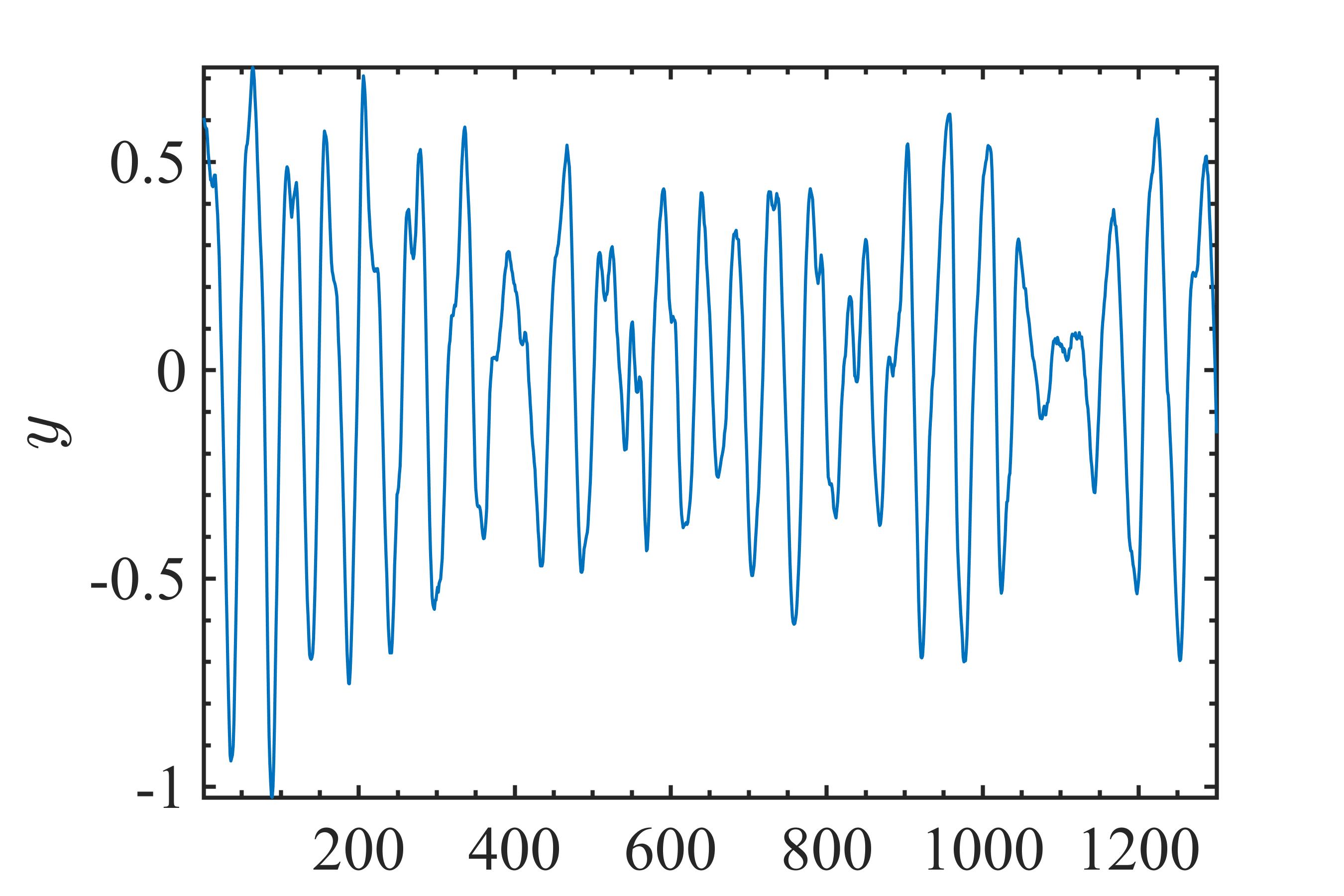}
  \caption{Water particle velocity (output), $y$}
  \label{f:ywav}
\end{subfigure}

\caption{Wave force data}
\label{f:w1}
\end{figure*}

In this section, a practical problem of non-linear wave force identification is considered. The objective is to investigate if the 2D-UPSO algorithm could identify a practical non-linear system. For this purpose, the velocity and force time histories of a fixed cylinder were obtained from The University of Salford. The input-output data are decimated by a factor of 2 and is shown in Fig.~\ref{f:w1}. Further information about this data set can be found in~\citep{Swain:Billings:Stansby:1998}. 
\begin{figure*}[!t]
\centering
\small
\begin{subfigure}{.5\textwidth}
  \centering
  \includegraphics[width=\textwidth]{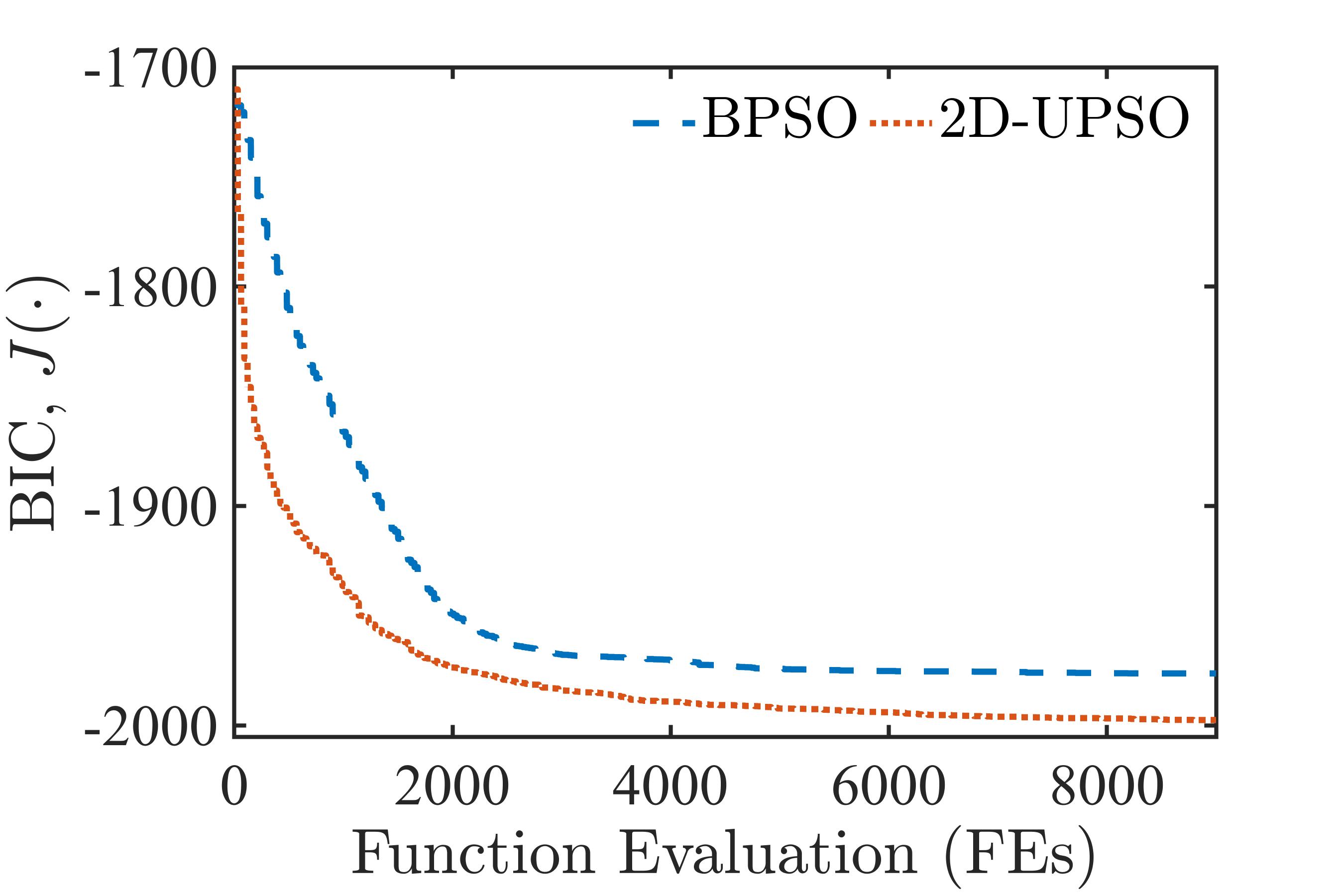}
  \caption{Criterion function, BIC $J(\cdotp)$}
  \label{f:jwav}
\end{subfigure}%
\hfill
\begin{subfigure}{.5\textwidth}
  \centering
  \includegraphics[width=\textwidth]{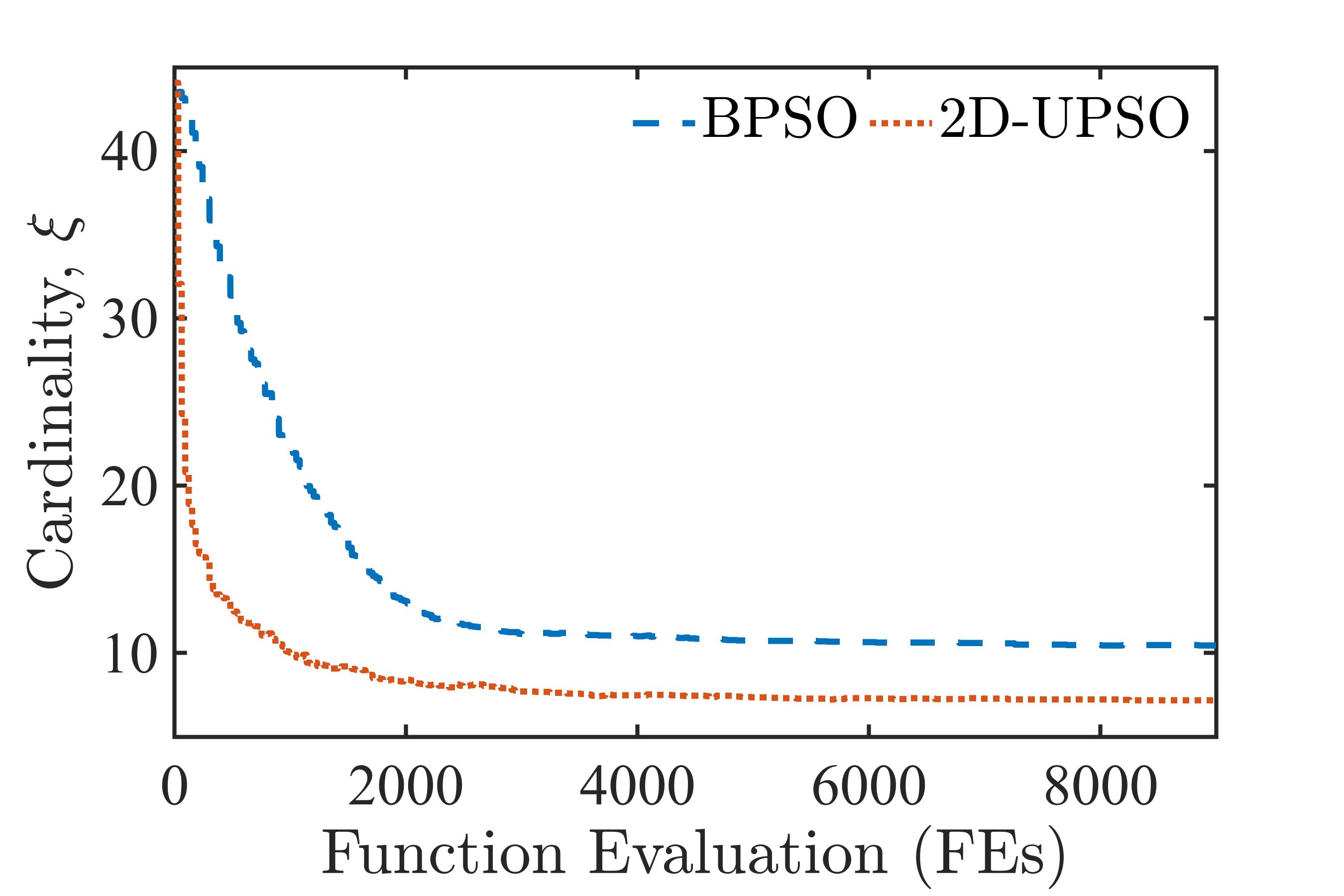}
  \caption{Cardinality, $\xi$}
  \label{f:xiwav}
\end{subfigure}

\caption{Average convergence plots over 40 independent runs of BPSO and 2D-UPSO on wave force data. a) Criterion function, BIC $J(\cdotp)$ b) Cardinality, $\xi$}
\label{f:w2}
\end{figure*}

The model is fitted between the in-line force (\textit{input}, `$u$') and the horizontal water particle velocity (\textit{output}, `$y$') considering $900$ data points and it is validated with a sequence of over $400$ points of input-output data taken arbitrarily from the rest of the available data points. Following similar procedure (Algorithm~\ref{fig:posprop}), polynomial NARX model is fitted using GA, BPSO and 2D-UPSO considering the initial NARX model set with $[n_u,n_y,n_l,N_t]=[7,7,3,680]$. 

Similar to the numerical examples considered in Section~\ref{s:resComp}, GA selected a large number of terms (\textit{approximately 50 terms in each run}). Further, the structures identified by GA provided high fitting-error (NMSE). Therefore, the results obtained by GA are not discussed here. The results obtained by BPSO and 2D-UPSO are shown in Fig.~\ref{f:w2}. These results show the dynamic search behavior averaged over 40 runs of the search algorithms, \textit{i.e.}, the variation in the criterion function, $J(\cdotp)$ and the cardinality of the best structure found hitherto as the search progresses. The results clearly indicate the benefits of the 2D learning; by exploiting the information about cardinality, 2D-UPSO could find comparatively smaller structures with a smaller fitting-error (NMSE) throughout the search.

The discrete NARX model identified by 2D-UPSO is given by, 
\begin{linenomath*}
\begin{small}
\begin{align}
\label{eq:eqsys}                               
    y(k) & = 1.3455 \, y(k-1) - 0.5964 \, y(k-4) + 0.19588 \, y(k-6) + 1.1545 \, u(k-1) \\
    &- 1.261 \, u(k-4) - 86.118 \, u(k-7)u(k-1)^2 + 88.116 \, u(k-4)^3\nonumber  
\end{align} 
\end{small}
\end{linenomath*}
\begin{figure*}[!t]
\centering
\small
\begin{subfigure}{.322\textwidth}
  \centering
  \includegraphics[width=\textwidth]{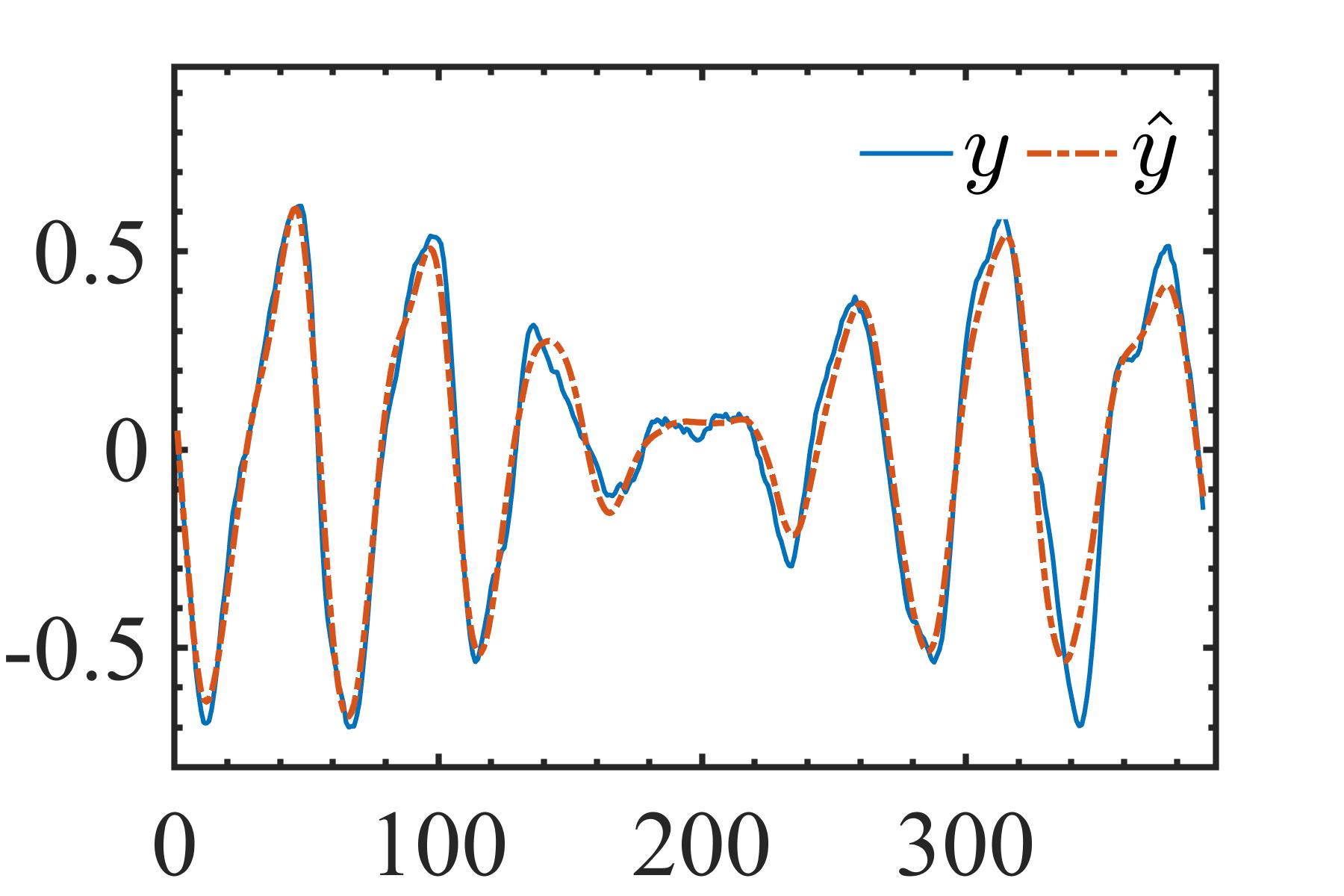}
  \caption{}
  \label{f:wavmpo}
\end{subfigure}%
\hfill
\begin{subfigure}{.322\textwidth}
  \centering
  \includegraphics[width=\textwidth]{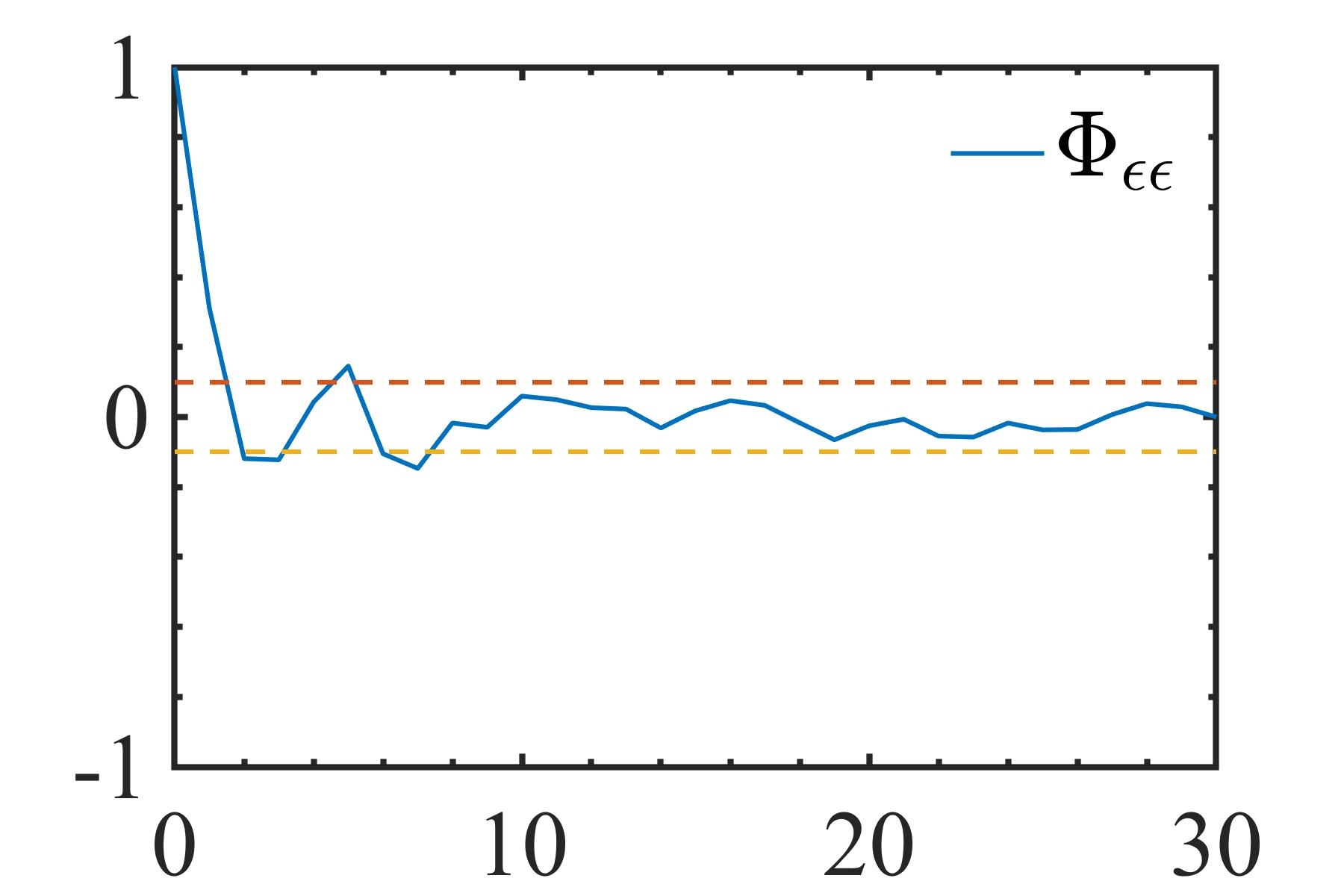}
  \caption{}
  \label{f:corr1}
\end{subfigure}
\begin{subfigure}{.322\textwidth}
  \centering
  \includegraphics[width=\textwidth]{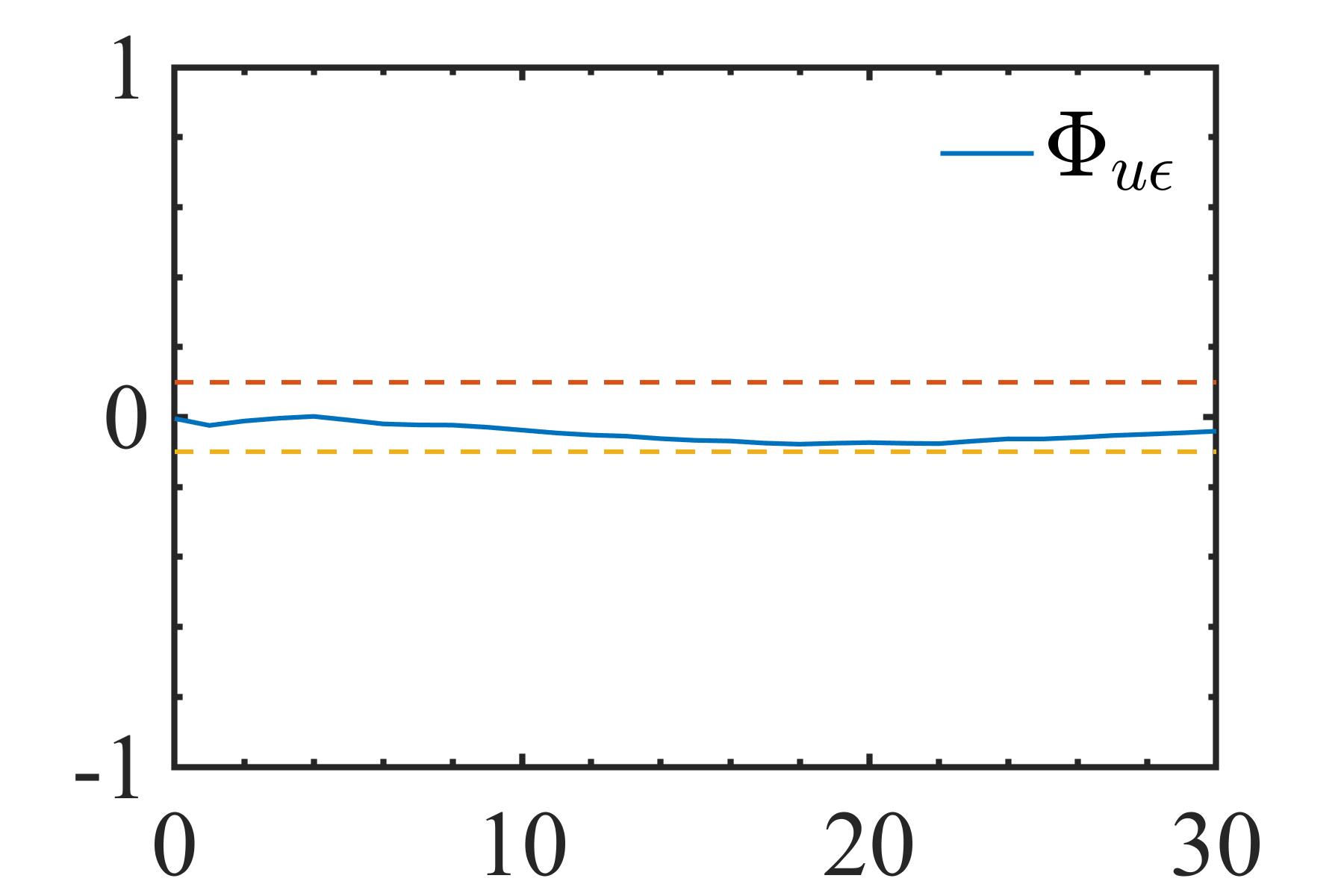}
  \caption{}
  \label{f:corr2}
\end{subfigure}%
\hfill
\begin{subfigure}{.322\textwidth}
  \centering
  \includegraphics[width=\textwidth]{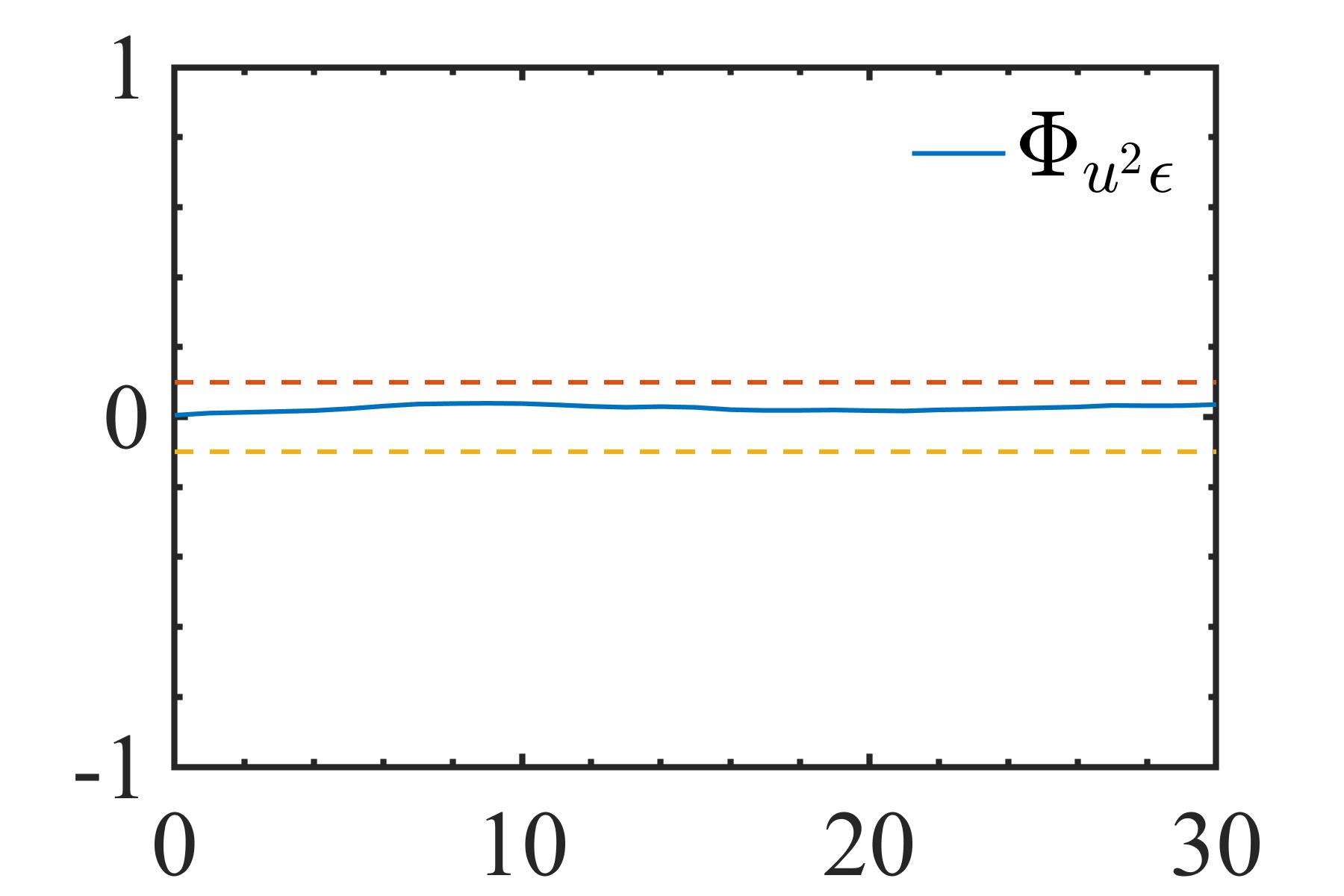}
  \caption{}
  \label{f:corr3}
\end{subfigure}
\begin{subfigure}{.322\textwidth}
  \centering
  \includegraphics[width=\textwidth]{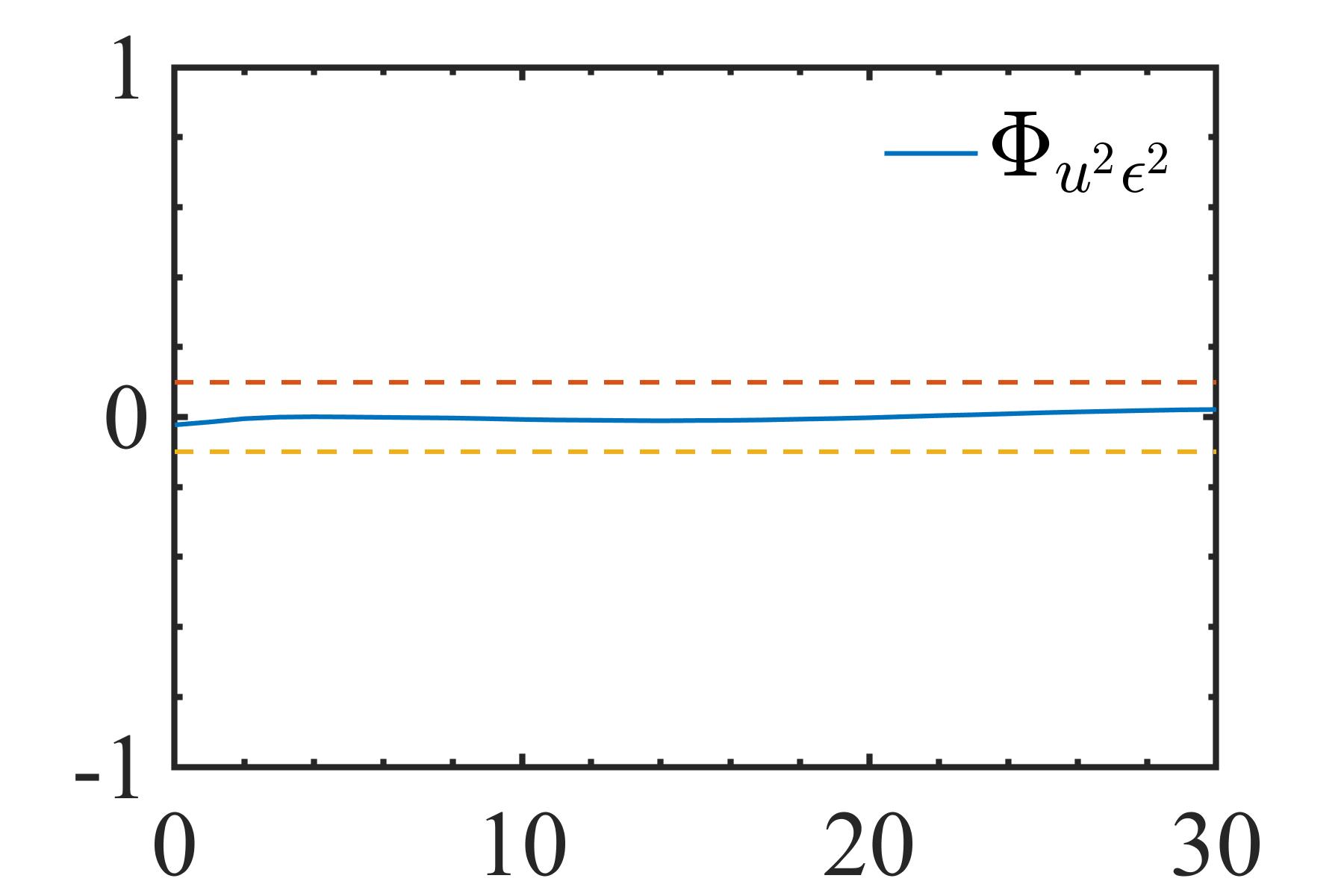}
  \caption{}
  \label{f:corr4}
\end{subfigure}%
\hfill
\begin{subfigure}{.322\textwidth}
  \centering
  \includegraphics[width=\textwidth]{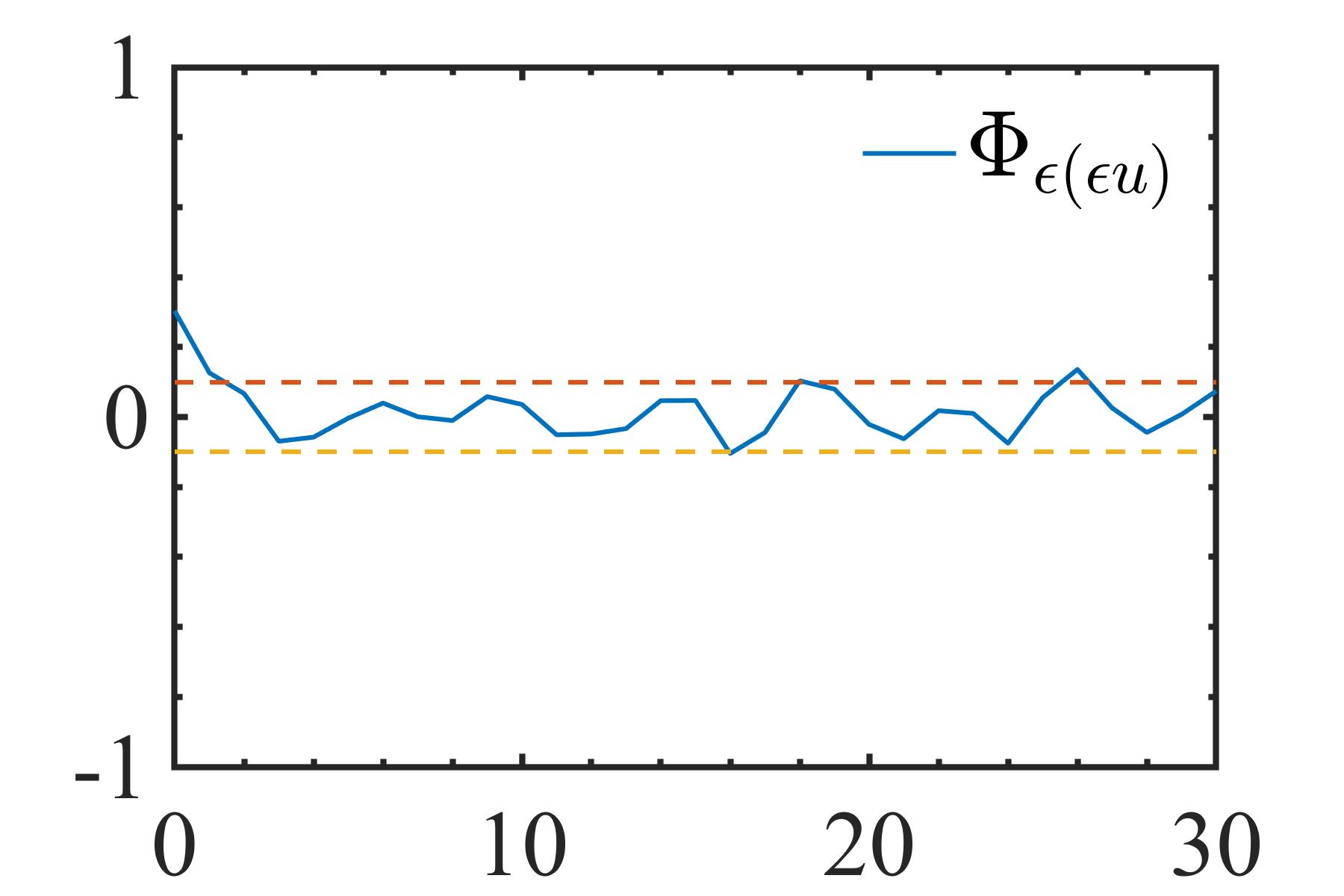}
  \caption{}
  \label{f:corr5}
\end{subfigure}

\caption{Model Validation. (a) Model predicted output. (b)-(f) Correlation tests}
\label{f:wavres}
\end{figure*}
The model is validated by computing the model predicted output over validation data and the cross-correlation plots are shown in Fig.~\ref{f:wavres}. From the results, it is observed that the identified NARX model could successfully encode the dynamics of the wave-force data. Note that although the identified model gave practically acceptable $NMSE$ of $2.12\times 10^{-1}$, further improvement in the model accuracy may be obtained by fitting a NARMAX model, which includes a noise model, to reduce/eliminate bias in the least square estimates. The proposed 2D-UPSO can be extended to identify NARMAX models which will be reported in the future.

\section{Conclusions}
\label{s:con}

A new Two-dimensional (2D) learning approach of selecting the correct structure of polynomial NARX models has been proposed. One of the key advantage of this algorithm is that the information about the \textit{number of terms} is explicitly incorporated into the search process. This property significantly reduces the number of spurious terms in the identified model. The effectiveness of the proposed approach has been demonstrated considering both simulated and practical non-linear systems. For discrete-time and practical nonlinear systems, the identified models are validated using the well-known cross-validation and correlation-based model validity tests. For continuous-time nonlinear systems, the identified discrete NARX models are further validated by comparing the generalized frequency response functions (GFRFs) which demonstrate that 2D-UPSO could successfully capture the system dynamics.

\linespread{1}

\appendix
\small
\section{Illustrative Example: Solution Representation}
\label{s:appsolr}
Consider a simple NARX model with a total of $5$ terms ($N_t=5$) as follows:
\begin{linenomath*}
\begin{align}
\label{eq:NARXExample}
	X_{model} & = \begin{bmatrix} x_1 & x_2 & x_3 & x_4 & x_5 \end{bmatrix} \\
			  & =\begin{bmatrix} y(k-1) & u(k-2) & y(k-3) & y(k-2)u(k-2) & u(k-3)^3 \end{bmatrix} \nonumber	  
\end{align}
\end{linenomath*}
For this problem, assume that the position of the $i^{th}$ particle is given by,
\begin{linenomath*}
\begin{align}
\label{eq:NARXExample1}
    \beta_i & = \begin{bmatrix} 1 & 0 & 1 & 1 & 0 \end{bmatrix}
\end{align}
\end{linenomath*}
This implies that only the \textit{first, third} and \textit{fourth} terms from the set $X_{model}$ are included into the \textit{structure/term subset}. Thus, the structure `$X_i$' encoded by the particle $\beta_i$ is given by,
\begin{linenomath*}
\begin{align}
\label{eq:NARXExample2}
    X_i & = \begin{bmatrix} x_1 & x_3 & x_4 \end{bmatrix} = \begin{bmatrix} y(k-1) & y(k-3) & y(k-2)u(k-2) \end{bmatrix}
\end{align}
\end{linenomath*}

\section{Illustrative Example: Evaluation of the Learning Sets}
\label{s:applearn}

Assume that the learning exemplar, $\alpha$, for the problem considered in~(\ref{eq:NARXExample}) is given by,
\begin{linenomath*}
\begin{align}
\label{eq:NARXExample3}
    \alpha & = \begin{bmatrix} 0 & 1 & 1 & 1 & 1 \end{bmatrix}
\end{align}
\end{linenomath*}
Following Algorithm~\ref{alg:learningset}, the learning sets are derived from $\alpha$ and $\beta_i$ as follows: 
\begin{enumerate}
    \item Set cardinality learning sets to null-vector (Algorithm~\ref{alg:learningset}, Line~\ref{line:ls1}):\\
    $\varphi_{\alpha} = \varphi_i = \begin{bmatrix} 0 & 0 & 0 & 0 & 0 \end{bmatrix}$
    \smallskip
    \smallskip
    \item Cardinality of $\alpha$ and $\beta_i$ (Algorithm~\ref{alg:learningset}, Line~\ref{line:ls2}):\\
    $\xi_{\alpha} = \sum \limits_{m=1}^{N_t} \alpha_m = 4$ and $\xi_i= \sum \limits_{m=1}^{N_t} \beta_{i,m} = 3$.
    \smallskip
    \smallskip
    \item Set the $\xi^{th}$ bit of $\varphi$ to `$1$' (Algorithm~\ref{alg:learningset}, Line~\ref{line:ls3}):\\
    $\varphi_{\alpha} = \begin{bmatrix} 0 & 0 & 0 & 1 & 0 \end{bmatrix}$ and $\varphi_i = \begin{bmatrix} 0 & 0 & 1 & 0 & 0 \end{bmatrix}$
    \smallskip
    \smallskip
    \item Evaluate feature learning sets (Algorithm~\ref{alg:learningset}, Line~\ref{line:ls4}):\\
    $\psi_{\alpha} = \{ \alpha \wedge \overline{\beta}_i \} = \begin{bmatrix} 0 & 1 & 0 & 0 & 1 \end{bmatrix}$ and $\psi_i = \beta_i = \begin{bmatrix} 1 & 0 & 1 & 1 & 0 \end{bmatrix}$
    \smallskip
    \smallskip
    \item Evaluate the final learning sets (Algorithm~\ref{alg:learningset}, Line~\ref{line:ls5}):\\
        $\mathcal{L}_{\alpha} = \begin{bmatrix} \varphi_{\alpha} \\ \psi_{\alpha} \end{bmatrix} = \begin{bmatrix} 0 & 0 & 0 & 1 & 0 \\ 0 & 1 & 0 & 0 & 1 \end{bmatrix}$ and $\mathcal{L}_{i} = \begin{bmatrix} \varphi_{i} \\ \psi_{i} \end{bmatrix} = \begin{bmatrix} 0 & 0 & 1 & 0 & 0 \\ 1 & 0 & 1 & 1 & 0 \end{bmatrix}$ 
\end{enumerate}

\section{Illustrative Example: Position Update}
\label{s:apppos}
Let the velocity of $i^{th}$ particle for this problem be given by,
\begin{linenomath*}
\begin{align}
\label{eq:example}
V_i & =\begin{bmatrix} v_{11}^i & v_{12}^i & v_{13}^i & v_{14}^i & v_{15}^i \\          
                       v_{21}^i & v_{22}^i & v_{23}^i & v_{24}^i & v_{25}^i \end{bmatrix} = \begin{bmatrix} 1.15 & 1.66 & 2.98 & 2.21 & 1.42\\
                       2.32 & 4.54 & 1.71 & 3.27 & 2.89 \end{bmatrix}
\end{align}
\end{linenomath*}

Following the Algorithm~\ref{fig:posprop}, the new position of the $i^{th}$ particle is determined as follows:
\begin{enumerate}
    \item Selection of the cardinality, (Algorithm~\ref{fig:posprop}, Line~\ref{line:pos1}-\ref{line:pos3}):
        \begin{itemize}
        \smallskip
            \item Selection likelihoods for \textit{cardinality}, \\
            $ \begin{bmatrix} v_{11}^i & v_{12}^i & \dots & v_{15}^i \end{bmatrix} = \begin{bmatrix} 1.15 & 1.66 & 2.98 & 2.21 & 1.42 \end{bmatrix}$
            \smallskip
            \item Cumulative likelihoods, $\Sigma = \begin{bmatrix} 1.15 & 2.81 & 5.79 & 8.0 & 9.42 \end{bmatrix}$
            \smallskip
            \item Selection probabilities, $ p = \begin{bmatrix} 0.12 & 0.30 & 0.61 & 0.85 & 1 \end{bmatrix}$
            \smallskip
            \item Let random number $r\in[0,1]$ equals to $0.4$. 
            \smallskip
            \item $ p_2 < r < p_3 $ which gives $\xi_i=3$. \\
            This implies that a total of $3$ number of terms to be included in the structure. The next step is to determine which $3$ terms should be included.
        \end{itemize}
        \smallskip
        \smallskip
    \item Selection of the terms (Algorithm~\ref{fig:posprop}, Line~\ref{line:fs1}-\ref{line:fs2}):
        \begin{itemize}
        \smallskip
            \item Selection likelihoods for the \textit{terms}, \\
            $ \begin{bmatrix} v_{21}^i & v_{22}^i & \dots & v_{25}^i \end{bmatrix} = \begin{bmatrix} 2.32 & 4.54 & 1.71 & 3.27 & 2.89 \end{bmatrix}$
            \smallskip
            \item Sort the term likelihoods in descending order and store the rankings in `$\tau$',\\
            $\tau = \begin{bmatrix} 4 & 1 & 5 & 2 & 3 \end{bmatrix}$
            \smallskip
            \item Selection of the terms as per $\xi_i$,\\
            $\xi_i=3$ and $\tau = \begin{bmatrix} 4 & 1 & 5 & 2 & 3 \end{bmatrix}$ which gives, $\beta_i= \begin{bmatrix} 0 & 1 & 0 & 1 & 1 \end{bmatrix}$
        \end{itemize}
\end{enumerate}
    
`$\beta_i$' gives the new position of the particle for the example considered in (\ref{eq:NARXExample}). Further, $\beta_i$ denotes that in the new structure all the terms are included except $x_1$ and $x_3$, \textit{i.e.},
\begin{linenomath*}
\begin{align*}
    X_i & = \begin{bmatrix} x_2 & x_4 & x_5 \end{bmatrix} = \begin{bmatrix} u(k-2) & y(k-2)u(k-2) & u(k-3)^3 \end{bmatrix} 
\end{align*}
\end{linenomath*}

\section{System Terms}
\label{s:app}
\begin{itemize}
\small
    \item System $S1$ : $X^{\star} = \{ x_1, \dots x_4\}$, where, \\
    \smallskip
    $x_1 \rightarrow y(k-1)$, $x_2 \rightarrow u(k-1)$, $x_3 \rightarrow u(k-1) \ast y(k-1)$ and $x_4 \rightarrow u(k-1)^2$
    \smallskip
    \item System $S2$ : $X^{\star} = \{ x_1, \dots x_5\}$, where,\\
    \smallskip
    $x_1 \rightarrow c$, $x_2 \rightarrow y(k-1)$, $x_3 \rightarrow u(k-2)$, $x_4 \rightarrow u(k-1)^2$ and $x_5 \rightarrow y(k-2)^2$
    \smallskip
    \item System $S3$ : $X^{\star} = \{ x_1, \dots x_4\}$, where,\\ \smallskip
    $x_1 \rightarrow y(k-1)$, $x_2 \rightarrow u(k-1)$, $x_3 \rightarrow u(k-1)^2$ and $x_4 \rightarrow u(k-1)^3$
    \smallskip%
    \item System $S4$ : $X^{\star} = \{ x_1, \dots x_{5}\}$, where,\\ \smallskip
    $x_1 \rightarrow y(k-1)$, $x_2 \rightarrow u(k-1)$, $x_3 \rightarrow y(k-2)^2$, $x_4 \rightarrow y(k-2) \ u(k-1)^2$, $x_5 \rightarrow u(k-3)^3$
    \smallskip
    \item System $S5$ : $X^{\star} = \{ x_1, \dots x_{4}\}$, where,\\ \smallskip
    $x_1 \rightarrow y(k-1) \ u(k-1)$, $x_2 \rightarrow y(k-2)$, $x_3 \rightarrow u(k-2)^2$, $x_4 \rightarrow y(k-2) \ u(k-2)^2$
    \smallskip
    \item System $S6$ : $X^{\star} = \{ x_1, \dots x_{4}\}$, where,\\ \smallskip
    $x_1 \rightarrow y(k-1)^3$, $x_2 \rightarrow y(k-1) \ u(k-1)$, $x_3 \rightarrow u(k-2)^2$, $x_4 \rightarrow y(k-2) \ u(k-2)^2$, $x_5 \rightarrow y(k-2)$

\end{itemize}

\bibliographystyle{elsarticle-num}


\end{document}